\documentstyle[11pt,epsfig]{article}
\newcommand{\be}{\begin{equation}}
\newcommand{\ee}{\end{equation}}
\newcommand{\bea}{\begin{eqnarray}}
\newcommand{\eea}{\end{eqnarray}}
\newcommand{\bc}{\begin{center}}
\newcommand{\ec}{\end{center}}
\newcommand{\vac}{\mid\! 0\rangle}
\newcommand{\vacl}{\langle 0\!\mid}
\newcommand{\x}{{\vec{\rm x}}}
\newcommand{\ka}{{\vec{\rm k}}}
\newcommand{\y}{{\vec{\rm y}}}
\newcommand{\q}{{\vec{\rm q}}}
\newcommand{\p}{{\vec{\rm p}}}

\newcommand{\rv}{{\vec{\rm r}}}

\newcommand{\s}{{\vec{\rm s}}}

\newcommand{\jv}{\vec{J}}
\newcommand{\kap}{\vec{\kappa}}
\newcommand{\di}{\displaystyle \int }
\newcommand{\PP}{{\cal P}}
\newcommand{\Pop}{{\mbox{\bf P}} }

\newcommand{\G}{{\mbox{\bf G}} }
\newcommand{\Aop}{{\mbox{\bf A}} }
\newcommand{\Bop}{{\mbox{\bf B}} }
\newcommand{\Cop}{{\mbox{\bf C}} }
\newcommand{\Qop}{{\mbox{\bf Q}} }
\renewcommand{\tilde}[1]{\widetilde{#1}}

\newcommand{\A}{\tilde{A}}
\newcommand{\B}{\tilde{B}}

\newcommand{\vc}{{\vec{\rm C}}}
\newcommand%
{\vf}%
[2]%
{\phi^{\pm (#1)}_{{\cal P}q}(#2)_{\alpha\beta}}%
\newcommand%
{\TT}%
[2]%
{{{\cal T}^{(#1)\pm}_{{\cal P}}(q;#2)}}%
\newcommand%
{\MV}%
[1]%
{{<\!\!\!<\!\!#1\!\!>\!\!\!>}}%
\def\bm#1{%
\mathchoice
{{\hbox{\boldmath$\displaystyle#1$\unboldmath}}}%
{{\hbox{\boldmath$\textstyle#1$\unboldmath}}}%
{{\hbox{\boldmath$\scriptstyle#1$\unboldmath}}}%
{{\hbox{\boldmath$\scriptscriptstyle#1$\unboldmath}}}%
}
\def\vec#1{\bm{#1}}

\title{Two- and Three-particle States in a Nonrelativistic Four-fermion
Model in the Fine-tuning Renormalization Scheme.
Goldstone mode "against" extension theory.
}

\author{
A.N.Vall, S.E.Korenblit, V.M.Leviant, D.V.Naumov, and A.V.Sinitskaya
}

\begin{document}
\maketitle

\bc
Irkutsk State University, 664003, Gagarin blrd, 20, Irkutsk, Russia
\footnote{E-mail KORENB@ic.isu.ru}
\ec
%%%%%%%%%%%%%%%%%%%%%%%%%%%%%%%%%%%%%%%%%%%%%%%%%%%%%%%%

\begin{abstract}
{In a nonrelativistic contact four-fermion model we show that simple
regularisation prescriptions together with a definite fine-tuning of the
cut-off-parameter dependence of ``bare'' quantities give the exact solutions
for the two-particle sector and Goldstone modes.
Their correspondence with the self-adjoint extension into Pontryagin space
is established leading to self-adjoint semi-bounded Hamiltonians in
three-particle sectors as well. Renormalized Faddeev equations for the
bound states with Fredholm properties are obtained and analysed.}
\end{abstract}

%%%%%%%%%%%%%%%%%%%%%%%%%%%%%%%%%%%%%%%%%%%%%%%%%%%%%%%%%%%%%%%%%%
\section{Introduction}
%%%%%%%%%%%%%%%%%%%%%%%%%%%%%%%%%%%%%%%%%%%%%%%%%

Models with four-fermion interactions arise in a wide range of problems
both in quantum field theory and condensed matter physics \cite{UMT}.
Four-fermion contact interaction models also shed light on the low-energy
hadronization regime of QCD where the perturbative approach fails.
They are used as qualitative and quantitative descriptions of various
phenomenological data in hadron physics. The non-perturbative nature of the
bound states in both hadron and condensed matter physics challenges numerous
efforts to develop non-perturbative methods, which particularly aim at an
explicit non-perturbative solution of the corresponding theoretical model
\cite{AMP}.

The success of four-fermion models originates, firstly, from the fact that
these models embody chiral symmetry and its spontaneous breaking \cite{Klev}.
It is well known, however, that such models are nonrenormalizable within
conventional perturbation theory. Calculations around four-fermion models
face ultraviolet divergencies. These divergences are treated, as a rule, by
introducing an ultraviolet cut-off $\Lambda$ indicating the range of
validity of the model.
The mathematical reason of the divergences partially become apparent in the
framework of extension theory. The very singular interactions in such models
cannot be considered as a correct quantum-mechanical potential. Therefore,
every $N-$ particle Fock state has to be studied within the prescriptions of
extension theory.

The nonrelativistic contact four-fermion models are particularly
interesting, because in these frameworks they possess a family of the
exact analytical three-dimensional solutions in the one- and two-particle
sectors. These solutions, for example, can be considered as a basis to study
the mechanism of bosonization and condensation in Hartree-Fock approximation.

It should be stressed that a vector current-current contact term leads to a
generalized point two-particle interaction which, in the modern extension
theory, appears simultaneously as a local and separable finite-rank
perturbation containing a finite set of arbitrary extension parameters with
clear physical meaning.  Thus, in contrast to some popular belief, the
contact field interaction promises to become physically even richer and more
predictive than the usual (non-local) separable one.

The nonrelativistic limit of the contact four-fermion model was developed
in our previous articles \cite{ALSHT}, \cite{YdFz}. There it was
demonstrated that such contact quantum field models possess exact
two-particle solutions. We clarified the mathematical origin of the
model divergences and gave a simple prescription how to treat them
nonperturbatively. To this end a functional dependence of all ``bare''
quantities on a cut-off $\Lambda$ was assumed. Next, this functional
dependence was determined by means of the limiting procedure relating the
finite observables and infinite ``bare'' quantities at
$\Lambda\rightarrow \infty$ in one-particle and two-particle Fock states.
In the present paper the investigation of our model is continued in order
to include the three-particle sector as well. It will be elucidated, how the
vacuum, one-particle, and two-particle renormalized Fock states completely
define the three-particle ones, demonstrating self-consistency of our
renormalization prescription, whose mathematical basis is provided by the
extension theory.

The paper is organized as follows. In section 2 the operator diagonalization
of the initial Hamiltonian is described. In section 3 and in Appendix A the
underlying singular two-particle problem is reviewed.
Sections 4, 5, and 6 contain our main analysis of three-particle
equations with some details placed in Appendix B. One can trace
the long history of the development of singular two- and
three-particle problems in the recent articles \cite{MMM} (and
references therein). We would like to notice here that our consideration
follows the idea of refs. \cite{BrFdd}, \cite{Thorn}, and especially
\cite{Ber}, but we use another possibility to regularize the instantaneous
(anti) commutation relations with the same regularization as for the
interaction.

\section{Contact Four-Fermion Models}

Let us consider the following four-fermion Hamiltonian
\be
H =\int d^3x\left\{\Psi^{\dagger a}_{\alpha }(x)\,{\cal E}({\Pop})\,
\Psi^a_{\alpha }(x) -\frac{\lambda}{4}\left[S^2(x)-
{\jv}^2(x)\right]\right\},
\label{H2}
\ee
where $x=(\x,t=x_0)$,
${\Pop}=-i\stackrel{\rightarrow}{\vec{\nabla}}{\!\!}_{\rm x}$,
$\stackrel{\longleftrightarrow}{\Pop}=
-i\left(\stackrel{\rightarrow}{\vec{\nabla}}{\!\!}_{\rm x}-
\stackrel{\leftarrow}{\vec{\nabla}}{\!\!}_{\rm x}\right)
$,
\be
S(x)=\Bigl(\Psi^{\dagger a}_{\alpha }(x)\,\Psi^a_{\alpha }(x)\Bigr),\quad
{\jv}(x)=(2mc)^{-1}\Bigl(\Psi^{\dagger a}_{\alpha }(x)
\stackrel{\longleftrightarrow}{{\Pop}}\Psi^a_{\alpha }(x)\Bigr),
\label{SJ}
\ee
with the fermion fields $\Psi^a_{\alpha }(x)$ satisfying the anticommutation
relations
\be
\left\{\Psi^a_{\alpha }(x)\,,\Psi^b_{\beta}(y)\right\}\Bigr|_{x_0=y_0}=0,
\;\;\left\{\Psi^a_{\alpha }(x)\,,\Psi^{\dagger b}_{\beta}(y)\right\}
\Bigr|_{x_0=y_0}=\delta_{\alpha\beta}\delta^{ab}\,\delta_3(\x-\y),
\label{ACR}
\ee
for $a,\alpha=1,2$ and with the convention
\be
\left\{\Psi^a_{\alpha }(x)\,,\Psi^{\dagger b}_{\beta}(y)\right\}
\Bigr|_{x_0=y_0;\x=\y}\longrightarrow
\delta_{\alpha\beta}\delta^{ab}\,\frac{1}{V^*}.
\label{conv}
\ee
Here ${\cal E}(k)$ is arbitrary ``bare'' one-particle spectrum, $V^*$ has
the meaning of an excitation volume and can be expressed through the usual
momentum cut-off parameter $\Lambda$.
The Hamiltonian is invariant under the (global) symmetry transformations
$SU_J(2)\times SU_T(2)\times U(1)$ generated by ($\sigma^i,\tau^r$ are Pauli
matrices)
\[
\left.\begin{array}{c}{\cal J}^i \\{\cal T}^r_{\alpha\beta}\end{array}\;
\right\}=
\frac{1}{2}\int d^3{\rm x}\,\Psi^{\dagger a}_\alpha (x)\left\{
\begin{array}{c}(\sigma^i)_{\alpha\beta}\delta^{ab} \\
(\tau^r)^{ab} \end{array}\right\}\Psi^b_\beta (x), \;\;{\cal U} =
\frac{1}{2}\int d^3{\rm x}\,\Psi^{\dagger a}_\alpha (x)\Psi^a_\alpha (x),
%\label{SUtj}
\]
where ${\cal J}^i$ are generators of "isotopic" $SU_J(2)$ transformations,
${\cal T}^r=\delta_{\alpha\beta}{\cal T}^r_{\alpha\beta}$ are generators
of additional - "colour" $SU_T(2)$ transformations and ${\cal U}$ is the
$U(1)$ charge. Such symmetry definitions are conditional. For example, one
can find the interaction structure (\ref{H2}) with the usual ${\cal J}$-spin,
as a direct nonrelativistic limit of the relativistic four-fermion
combination
$(\overline{\psi^a}\psi^a)^2+(\overline{\psi^a}\gamma^{\nu}\psi^a)^2$,
neglecting the magnetization current $\vec{\nabla}\times\Bigl(
\Psi^{\dagger a}_{\alpha}(x)\vec{\sigma}_{\alpha\beta}
\Psi^a_{\beta}(x)\Bigr)$ in comparison with ${\jv}(x)$, i.e.
eliminating usual spin-orbit and spin-spin interactions.
This elimination is coordinated with our subsequent consideration.

Introducing Heisenberg fields in a momentum representation
\be
\Psi^a_{\alpha}(\x,t)=\int\frac{d^3k}{(2\pi)^{3/2}}\,e^{i(\ka\x)}\,
b^a_{\alpha}(\ka,t),\;\;
\left\{b^a_{\alpha}(\ka,t)\,,b^{\dagger b}_{\beta}(\q,t)\right\}=
\delta_{\alpha\beta}\delta_{ab}\delta_3(\ka-\q),
\label{HF}
\ee
we consider at $t=0$ their three different linear operator realizations
via physical fields by Bogoliubov rotations with $u_a=\cos\theta^a,\
v_a=\sin\theta^a $ and purely antisymmetric $\epsilon_{\alpha\beta}$,
$\epsilon_{\alpha\gamma}\epsilon_{\beta\gamma}=\delta_{\alpha\beta}$:
\bea
&& b^a_{\alpha}(\ka,0)=e^{\G}\,d^a_{\alpha}(\ka)\,e^{-\G}=
u_a\,d^a_{\alpha}(\ka)-v_a\epsilon_{\alpha\beta}d^{\dagger a}_{\beta}(-\ka),
\nonumber \\
&& \G=\frac{1}{2}\sum_{a=1,2}\theta^a\epsilon_{\alpha\beta}\int d^3k
\left[d^{\dagger a}_{\alpha}(\ka)d^{\dagger a}_{\beta}(-\ka)+
d^a_{\alpha}(\ka)d^a_{\beta}(-\ka)\right]=-\G^{\dagger}.
\nonumber
\eea
Under the condition $u_a v_a=0$ for $a=1,2$ this gives some reduced
Hamiltonians in normal form which are {\it exactly diagonalizable} on the
suitable vacua:
\be
H={\cal V}w_0+\widehat{H},\qquad \widehat{H}=\widehat{H}_0+\widehat{H}_I,
\label{H+H}
\ee
with ${\cal V}$- being a space volume,
\bea
&&w_0=\frac{1}{V^*}\left[\left(2\MV{{\cal E}(k)}-4g\right)
\left(v_1^2+v_2^2\right)-8g (v_1 v_2)^2\right],
%\label{aW0} \\
\nonumber \\
&& \MV{{\cal E}(k)}\stackrel{def}{=}V^*\int \frac{d^3k}{(2\pi)^3}
\,{\cal E}(k),\;\;\frac{1}{V^*}=\frac{\Lambda^3}{6\pi^2},\;\;
\MV{k^2}\,=\frac{3}{5}\Lambda^2,\;\; g=\frac{\lambda}{4V^*},
\label{k2} \\
&&d^a_{\alpha}(\ka)\vac=0,\;\;\widehat{H}\{d\}\vac =0,\;\;
\left[\widehat{H}\{d\}\,,d^{\dagger a}_{\alpha}(\ka)
\right]\vac =E^a(k)\,d^{\dagger a}_{\alpha}(\ka)\vac ,\quad
\label{dVac} \\
&& E^a(k)=\frac{g}{(2mc)^2}\left(k^2+\MV{k^2}\right)+g+
(1-2v_a^2)\left[{\cal E}(k)-2g(1+2v_{3-a}^2) \right],\;
\label{Eav} \\
&& \widehat{H}_0\{d\}=\sum_{a=1,2}\di d^3k\, E^a(k)\,d^{\dagger a}_{\alpha}(\ka)
\,d^a_{\alpha}(\ka),
\label{aH0} \\
&&\widehat{H}_I\{d\}=\sum_{a,b}\di d^3k_1d^3k_2d^3k_3
d^3k_4\;\delta(\ka_1+\ka_2-\ka_3-\ka_4)
\nonumber \\
&&\cdot K^{(ab)}\left(\frac{\ka_1-\ka_2}{2};\frac{\ka_4-\ka_3}{2}\right)
d^{\dagger a}_{\alpha}(\ka_1)d^{\dagger b}_{\beta}(\ka_2)d^b_{\beta}(\ka_3)
d^a_{\alpha}(\ka_4),
\label{aHI}
\eea
in contrast to variational solutions with $u_a v_a\neq 0$, usually exploring
in the theory of superconductivity.
The different realizations correspond to different systems when $v_{1,2}$
independently take the values 0,1. For convenience we call them
A,B,C systems. \\
For the B-system: $v_1=v_2=0$, $\Bop_{\alpha}(\ka)=d^1_{\alpha}(\ka)$,
$\widetilde{\Bop}_{\alpha}(\ka)=d^2_{\alpha}(\ka)$, therefore
$E_B(k)=E^{1,2}_B(k)$.
One can see that the respective vacuum state $\vac_B$ is a singlet for both
the $SU_J(2)$ and $SU_T(2)$ groups and the one-particle excitations
of $B$ and $\B$ form the corresponding fundamental representations. \\
For the C-system: $v_1=v_2=1$,
$\epsilon_{\alpha\beta}\Cop_{\beta}(\ka)=d^1_{\alpha}(\ka)$,
$\epsilon_{\alpha\beta}\widetilde{\Cop}_{\beta}(\ka)=d^2_{\alpha}(\ka)$,
$E_C(k)=E^{1,2}_C(k)$.
The symmetry of this system is similar to the symmetry of the B-system. \\
For the A-system: $v_1=0$, $v_2=1$ (or $v_1=1$, $v_2=0$); it will be
considered in detail below. Let $\Aop_{\alpha}(\ka)=d^1_{\alpha}(\ka)$,
$\epsilon_{\alpha\beta}\widetilde{\Aop}_{\beta}(\ka)=d^2_{\alpha}(\ka)$,
and let $f^{ab}$ be an arbitrary constant $SU_T(2)$ matrix, then for
$ E^{(+,-)}_A(k)\equiv E_{\A,A}(k)\equiv E^{2,1}_A(k) $ the corresponding
Heisenberg fields (\ref{HF}) read (hereafter we write $\vac\equiv\vac_A$):
\bea
\left\{\!\!\!\begin{array}{c}
{\Aop}_{\alpha}(\ka,t)\, e^{-itE^{(-)}_A(k)} \\
\widetilde{\Aop}_{\alpha}(\ka,t)\, e^{-itE^{(+)}_A(k)}\end{array}\!\!\!
\right\}=e^{iHt}\left\{\!\!\!\begin{array}{c}\Aop_{\alpha}(\ka)\\
\widetilde{\Aop}_{\alpha}(\ka)\end{array}\!\!\!\right\}e^{-iHt},\;
\begin{array}{c}
{\Aop}_{\alpha}(\ka,t)\vac =\Aop_{\alpha}(\ka)\vac \\
{\Aop}^{\dagger}_{\alpha}(\ka,t)\vac =\Aop^{\dagger}_{\alpha}(\ka)\vac ,
\end{array}
\label{GFA} \\
\Psi^a_{\alpha}(x)_A=\!\int\!\frac{d^3k}{(2\pi)^{3/2}}
\left[f^{a1}{\Aop}_{\alpha}(\ka,t)e^{-itE^{(-)}_A(k)}+
f^{a2}\widetilde{\Aop}^{\dagger}_{\alpha}(-\ka,t)e^{itE^{(+)}_A(-k)}\right]\!
e^{i(\ka\x)}.
\nonumber
\eea
It is easy to show that for this A system the symmetries $SU_T(2)$ and
$U(1)$ turn out to be spontaneously broken and there are four composite
Goldstone states associating with spin-flip waves of vacuum "medium" --
possessed spontaneous "colour" magnetization $-{\cal L}$ in the
$\vec{n}$-direction \cite{UMT}. They are created by the operators
\cite{YdFz}
\bea
{\cal G}^{\pm}_{\alpha\beta}(0)={\cal T}^r_{\alpha\beta}\cdot
\frac{1}{2}\,{\rm Tr}\left\{\widehat{\vec{\rm f}}(\vec{n})\tau^{\pm}
\widehat{\vec{\rm f}}^{\dagger}(\vec{n})\tau^r\right\}=
\int d^3k \left\{\begin{array}{c}
\Aop^{\dagger}_\alpha(\ka)\,\widetilde{\Aop}^{\dagger}_{\beta}(-\ka) \\
\widetilde{\Aop}_\alpha(\ka)\,\Aop_{\beta}(-\ka) \end{array}\right\},\quad
\label{Gld0}
\eea
for which:
\bea
[{H}\,,{\cal G}^{\pm}_{\alpha \beta}(0)] = 0,\quad
\vacl{\rm T}\!\left(\Psi^a_{\alpha}(x)\vec{\tau}_{ba}
\Psi^{\dagger b}_{\beta}(x)\right)\!\vac =\delta_{\alpha\beta}\,
\frac{\vec{n}}{V^*},\quad
\label{cond} \\
-\,\vacl\left(\vec{n}\cdot\vec{\cal T}\right)\vac=
-\,\vacl\delta_{\alpha\beta}\,{\cal G}^{3}_{\alpha\beta}(0)\vac=
\vacl{\cal U}\vac =\frac{{\cal V}}{V^*}\equiv\,{\cal L},\qquad
\nonumber
\eea
because $f^{ab}=f^{ab}(\vec{n})$ in fact parametrizes some rotation
from z-direction to the $\vec{n}(\vartheta,\varphi)$-direction:
$f^{ab}(\vec{n})=e^{-i\varphi\rm T_3}e^{-i\vartheta\rm T_2}$, where
$\vec{\rm T}=\vec{\tau}/2$, $\tau^{\pm}=\tau^1\pm i\tau^2 $.

\section{Two-Particle Eigenvalue Problems}

The interaction between all particles in the systems B and C is the same,
as in the $AA$ and $\A\A$-channels of system A. So it is enough to
consider the last one. Hereafter $BB$ means $BB$, $\B\B$, $B\B$, and the
same for $CC$. Let us introduce the two-particle interaction kernels
occurring in Eq. (\ref{aHI}) and the two-particle energies as:
\bea
&& K^{(QQ')}(\s,\ka)=K^{\PP}_{\{\pm\}}(\s,\ka),
\nonumber \\
&&\mbox{ for } QQ'=
\left\{\tiny\begin{array}{c}\!\!\!\A\A,AA,BB,CC \\ A\A\!\!\!\end{array}
\right\}\mbox{ respectively,}
\nonumber \\
&& -2K^{\PP}_{\{\pm\}}(\s,\ka)=\frac{V^*}{(2\pi)^3}\cdot\frac{2g}{(2mc)^2}
\left[(\s+\ka)^2+Z^{\PP}_{\{\pm\}}\right],
\label{K2} \\
&& Z^{\PP}_{\{\pm\}}=\{\pm\}(2mc)^2-\PP^2,\qquad \label{L} \\
&&E^{QQ'}_2(\PP,\ka)= E_{Q}\left(\frac{\PP}{2}+\ka\right)+
E_{Q'}\left(\frac{\PP}{2}-\ka\right)= E_{2\{\pm \}}(\PP,\ka),\quad
\label{E2+} \\
&& E_{2\{\pm\}}(\PP,\ka)\equiv\frac{2g}{(2mc)^2}\left[\MV{k^2}+
Z^{\PP}_{\{\pm\}}+k^2+\frac{5}{4}\,\PP^2\right]+\Theta_{\{\pm\}},
\nonumber
\eea
with
\bea
&&\Theta_{\{+\}}=
\left\{\!\!\!\begin{array}{c}\displaystyle \!\!\!\!\!\!\pm\!
\left[4g-{\cal E}(\frac{\PP}{2}+\ka)-{\cal E}(\frac{\PP}{2}-\ka)\right] \\
\displaystyle \;
\pm\left[12g-{\cal E}(\frac{\PP}{2}+\ka)-{\cal E}(\frac{\PP}{2}-\ka)\right]
\end{array}\!\!\!\! \right\},\;\mbox{ for }\;\;
QQ'=\left\{\!\!\!\begin{array}{c}\A\A\\BB\\CC\\AA\end{array}\!\!\!\right\},
\nonumber \\
&&\Theta_{\{-\}}=
\left[{\cal E}(\frac{\PP}{2}+\ka)-{\cal E}(\frac{\PP}{2}-\ka)\right],\qquad
\quad \mbox{ for }\;\; QQ'=A\A.
\nonumber
\eea
Now we can formulate two-particle eigenvalue problems in the Fock eigenspace
of the kinetic part $\widehat{H}_0$ of the reduced Hamiltonian
$\widehat{H}$ in Eq. (\ref{H+H}):
\bea
&& \widehat{H}\mid R^{\pm (QQ')}_{\alpha\beta}(\PP,\q)\rangle =
E^{QQ'}_2(\PP,\q)\,\mid R^{\pm (QQ')}_{\alpha\beta}(\PP,\q)\rangle,
\label{egnvl} \\
&& \widehat{H}\mid {\rm B}^{\PP(QQ')}_{\alpha \beta} \rangle  =
M^{QQ'}_2(\PP)\mid {\rm B}^{\PP(QQ')}_{\alpha \beta} \rangle ,\;\;\;\;
M^{QQ'}_2(\PP)=E^{QQ'}_2(\PP,q=ib),
\nonumber \\
&& \mid R^{\pm (QQ')}_{\alpha\beta}(\PP,\q)\rangle =\di d^3k\,
\Phi^{\pm QQ'}_{\PP q}(\ka)\mid R^{0(QQ')}_{\alpha\beta}(\PP,\ka)\rangle,
\label{egnst} \\
&& \mid {\rm B}^{\PP(QQ')}_{\alpha \beta}\rangle = \int d^3k\,
\Phi^{QQ'}_{\PP b}(\ka)\mid R^{0(QQ')}_{\alpha\beta}(\PP,\ka)\rangle,
\label{egnb}  \\
&& \mid R^{0(QQ')}_{\alpha\beta}(\PP,\ka)\rangle =
\Qop^{\dagger}_{\alpha }(\frac{\PP}{2}+\ka)\,
\Qop'^{\dagger}_{\beta }(\frac{\PP}{2}-\ka) \vac ,
\nonumber
\eea
($\Qop,\Qop'$ stands for the creation operators $\Aop,\widetilde{\Aop}$, or
$\Bop,\widetilde{\Bop}$, or $\Cop,\widetilde{\Cop}$) in terms of the
Schr\"o\-din\-ger equation for the respective scattering or bound-state wave
functions:
\be
\left[E^{QQ'}_2(\PP,\ka)-M^{QQ'}_2(\PP)\right]\Phi^{QQ'}_{\PP b}(\ka)=
-2\int d^3s\,\Phi^{QQ'}_{\PP b}(\s)\, K^{(QQ')}(\s,\ka).
\label{Prblm}
\ee
It is easy to check \cite{ALSHT}, \cite{YdFz}, using for divergent integral
{\it the same} $\Lambda$-cut-off as for the definitions (\ref{k2}),
(\ref{Eav}), (\ref{E2+}) that at $m(\Lambda)\rightarrow\infty$
with $\Lambda\rightarrow\infty$ this equation for the case $\{-\}$, almost
independently of the very form of the ``bare'' spectrum ${\cal E}(k)$,
admits a simple solution
\be
\Phi^{A\A}_{\PP b}(\ka)= const,\quad M^{A\A}_2(\PP)=\frac{5}{4}
\frac{\PP^2}{{\cal M}_0},
\quad {\cal M}_0=\lim_{\Lambda\rightarrow\infty}\frac{(2mc)^2}{2g}.
\label{Gld}
\ee
It presents four Goldstone states in motion whose creation operators
${\cal G}^{+}_{\alpha \beta}(\PP)$ are defined by Eq. (\ref{egnb}).
For $\PP=0$ they are given by Eq. (\ref{Gld0}) and exactly commute with the
Hamiltonian (\ref{H2}). Thus, Eq. (\ref{Prblm}) holds true for $\PP=0$ with
the finite $\Lambda$ as well.
The conditions are required for $\PP\neq 0$ only:
\[{\cal E}(k)=mc^2 h(z^2),\;\;z=\frac{k}{mc},\;\;h'(0)<\infty ,\;\;
\lim_{k\rightarrow\infty}\left[{\cal E}(\frac{\PP}{2}+\ka)-
{\cal E}(\frac{\PP}{2}-\ka)\right]\cdot k^{-2}=0.
\]
It is worth to emphasize that this, in a certain sense, generalized solution
comes up only in an $A\A$-channel and that the Goldstone states remain
motionless without a vec\-tor-current contribution ${\jv}(x)$ in Eq.
(\ref{H2}), i.e. for $c=\infty$.

According to Eq. (\ref{Eav}), a quadratic form of the ``bare'' spectrum
transforms to the following renormalized one:
\bea
{\cal E}(k)=\frac{k^2}{2m}+{\cal E}_0\,\longmapsto\,E^{(\pm)}_Q(k)=
\frac{k^2}{2{\cal M}^{(\pm)}}+E^{(\pm)}_{Q0},\qquad\qquad\qquad
\label{EQ1} \\
E^{(\pm)}_{Q0}=g\left(\frac{\MV{k^2}}{(2mc)^2}+c^{(\pm)}_Q\right)\mp
{\cal E}_0,\;\;c^{(\pm)}_A\equiv c_{\left\{\!\!\!\tiny\begin{array}{c}
\A \\ A \end{array}\!\!\!\right\}}= -1\pm 4,\;\;c_{\left\{\!\!\!\tiny
\begin{array}{c}C(+)\\B(-)\end{array}\!\!\!\right\}}=3\pm 4,
\nonumber\\
\frac{1}{2{\cal M}^{(\pm)}}=\frac{g}{(2mc)^2}\mp \frac{1}{2m},\qquad
\lambda_0=\frac{\lambda {\cal M}^{(\pm)}}{2},\quad
\mu_0=\frac{\lambda_0}{(2mc)^2}.\qquad
\label{mug}
\eea
For both cases $\{\pm\}$ in Eqs. (\ref{K2}) and (\ref{E2+}) Eq.
(\ref{Prblm}) reveals in configuration space a strongly singular
point-interaction potential with the result:
\[\left(- \nabla^2_{\rm x}-q^2\right)\psi_q(\x)=\delta_3(\x)N_1(q)-
\nabla^2_{\rm x}\delta_3(\x)N_2(q)-2\mu_0
\left((\vec{\nabla}\psi_q)(0)\cdot\vec{\nabla}_{\rm x}\delta_3(\x)\right),
\]
\be
N_1(q)\equiv (\{\pm\}\lambda_0-\mu_0\PP^2)\psi_q(0)-
\mu_0(\nabla^2\psi_q)(0), \;\;\;\;N_2(q)\equiv\mu_0\psi_q(0).
\label{R12}
\ee
It was studied in refs. \cite{YShr}-\cite{VDj}.
The first and second terms on the r.h.s. of this equation represent an
interaction with angular momentum $l=0$, the third one gives an
interaction for $l=1$ only. Among the various solutions obtained in refs.
\cite{ALSHT} and \cite{YdFz} for the two-particle wave function of Eqs.
(\ref{egnst}) and (\ref{egnb}) that are induced by the various self-adjoint
extensions of a singular operator from (\ref{R12}) the use of the
$\Lambda$-cut-off regularization \cite{BrFdd} together with the simple
subtraction procedure for $\Lambda\rightarrow\infty$, picks out
(analogously to refs. \cite{Ber} and \cite{Shond3}) the following
renormalized solution (with the symbol $\Longrightarrow$ meaning
"is reduced to"):
\be
\mid l,{\rm J,m};\PP,q\rangle^{\pm}=\di d^3k\,\vf{l,\rm J,m}{\ka}\,
\mid R^{0(QQ')}_{\alpha\beta}(\PP,\ka )\rangle,
\label{2states}
\ee
where:
\bea
\vf{l,\rm J,m}{\ka}=\chi^{(\rm J,m)}_{\alpha \beta}\,
\Phi^{\pm(l)}_{{\cal P}q}(\ka),\quad\;\chi^{(1,\pm 1)}_{\alpha \beta}=
\left\{\begin{array}{c}\delta_{\alpha 1}\delta_{\beta 1}\\
\delta_{\alpha 2}\delta_{\beta 2} \end{array}\right.,
\quad\qquad\qquad
\label{phichi} \\
\chi^{(0,0)}_{\alpha \beta}=\frac{1}{\sqrt{2}}
(\delta_{\alpha 1}\delta_{\beta 2}-\delta_{\alpha 2}\delta_{\beta 1}),\quad
\chi^{(1,0)}_{\alpha \beta}=\frac{1}{\sqrt{2}}
(\delta_{\alpha 1}\delta_{\beta 2}+\delta_{\alpha 2}\delta_{\beta 1}),
\qquad\qquad
\label{SpSt} \\
\Phi^{\pm(l)}_{{\cal P}q}(\ka)=\frac{1}{2}\left[\delta_3(\ka-\q)+
(-1)^{l} \delta_3(\ka+\q)\right]+\frac{\TT{l}{k} }{k^2-q^2\mp i0 },
\qquad\quad
\label{Phil} \\
\TT{0}{k}=\left.\gamma V^*\frac{\gamma\MV{[k^2]}+Z^{\PP}+q^2+
(1-\gamma)k^2}{(2\pi)^3\left[{\cal D}^{\PP}(\mp iq)-{\cal D}^{\PP}(b)\right]}
\right|_{\Lambda\rightarrow\infty}\!\Longrightarrow\frac{T(\mp iq)}{2\pi^2}=
\nonumber \\
=\frac{-\Upsilon}{2\pi^2(\Upsilon+b\mp iq)(b\pm iq)},\quad
\left(Z^{\PP}=Z^{\PP}_{\{\pm\}}\right),\quad\Phi^{(0)}_{\PP b}(\ka)=
\frac{const}{(k^2+b^2)},\quad
\label{T0rnrm} \\
\TT{1}{k}=\left.2(\ka\cdot\q)\,\frac{\gamma V^*}{(2\pi)^3}\left[1-
\frac{2}{3}{\cal J}_1(\mp iq)\right]^{-1}\right|_{\Lambda\rightarrow\infty}
\Longrightarrow 0.\qquad\qquad
\label{T1}\\
\mbox{For }\;Q=Q':\;
\vf{l,\rm J,m}{\ka}=-\phi^{\pm (l,{\rm J,m})}_{{\cal P}q}
(-\ka)_{\beta\alpha},\qquad\;l={\rm J}=0,1.\quad
\label{QQ} \\
\mbox{Here we have: }\;g=\Lambda^2G(\Lambda),\;\;\;(2mc)^2=\Lambda^2\nu(\Lambda),
\;\;\;{\cal E}_0=\Lambda^2\epsilon(\Lambda),\qquad\qquad
\label{gamma1} \\
\gamma\equiv\gamma_{\{\pm\}},\;\;\;\gamma^{(\pm)}_{\{+\}}(\Lambda)=
\frac{2{\cal M}^{(\pm)}g}{(2mc)^2}=\frac{\mu_0}{V^*}=
1\pm\frac{{\cal M}^{(\pm)}}{m},\;\;\;\gamma_{\{-\}}\equiv 1,\qquad\quad
\label{gamma} \\
G(\Lambda)=G_0+G_1/\Lambda+G_2/\Lambda^2+\ldots,\qquad\qquad\qquad\quad
\label{LSer}
\eea
and similarly for $\nu(\Lambda),\,\epsilon(\Lambda),\,\gamma(\Lambda)$.
Thus, if $G_0,\nu_0\neq 0$, then one has
\bea
\gamma^{(\pm)}_{0\{+\}}=1,\;\;\;\gamma^{(\pm)}_{1\{+\}}=
\pm c\sqrt{\nu_0}/G_0,\qquad\qquad\qquad\qquad\qquad\qquad\qquad
\label{ft+} \\
{\cal M}^{(\pm)}_{0\{\pm\}}={\cal M}_0=\frac{\nu_0}{2G_0},\;\;\;
\Upsilon=\frac{\pi}{2}\left(\frac{3}{5}\gamma_1+\sigma+\nu_1\right),\;\;\;
\nu_{0\{\pm\}}=-\,\{\pm\}\frac{3}{5}.\qquad
\label{01}
\eea
The quantities ${\cal J}_n(\varrho)$ and ${\cal D}^{\PP}(\varrho)$ are
defined in Appendix A by Eqs. (\ref{Jnk2}) and (\ref{DDr}).
The Galilei invariance of this solution is restored only due to the limit
${\Lambda\rightarrow\infty}$ in the same manner as for the Goldstone
states above.
We notice from Eqs. (\ref{k2}), (\ref{T0rnrm}), an (\ref{gamma1}) that there
is no direct relation between the character of the point interaction and the
sign of the quartic contact self-interaction in Eq. (\ref{H2}).
One can always choose for a given $g(\Lambda)$ the $\Lambda$-dependence of
the "bare" parameters $m(\Lambda)$ and ${\cal E}_0(\Lambda)$ to leave the
${\cal M}(\Lambda)$ and $E_0(\Lambda)$ finite for $\Lambda\rightarrow\infty$.
On the contrary $g(\Lambda)$ is determined by the two-particle eigenvalue
problem. So, the last equality in Eq. (\ref{01}) reflects condition for the
existence of the bound state defined by Eqs. (\ref{DDr}) and (\ref{D02}),
which serves here as a dimension transmutation condition \cite{BrFdd},
\cite{Thorn} transforming the ``bare'' coupling constants $\lambda_0$ and
$\mu_0$ of Eq. (\ref{mug}) and the cut-off $\Lambda$ into unknown binding
and scattering dimensional parameters $b$ and $\Upsilon$ \cite{YdFz}.
In this way, these real quantities become {\it arbitrary} parameters of
the self-adjoint extension and some of them are expressed through the
coefficients of the formal $\Lambda$-series (\ref{LSer}) of ``bare''
quantities (\ref{gamma1}) by the fine-tuning relations (\ref{01}).
Within these relations the finite one-particle spectra for $QQ$-channels
take the following forms (the index columns in the l.h.s. being in direct
correspondence with the terms on the r.h.s.)
$$
E_{\{+\}}(k)\equiv
E_{\left\{\!\!\!\tiny\begin{array}{c} \A \\ B \\ C \\ A \end{array}\!\!\!
\right\}}(k)=\frac{k^2+\nu_2}{2{\cal M}_0}-\frac{5}{3}\nu_1
\left(G_1-\frac{\nu_1}{2{\cal M}_0}\right)+\left\{\begin{array}{c}
\displaystyle \pm(2G_2-\epsilon_2) \\ \displaystyle
\pm(6G_2-\epsilon_2) \end{array}\right\},
$$
for:
\be
\left\{\begin{array}{c}\displaystyle
\left(\epsilon_0=2G_0,\;\;\epsilon_1=2G_1\pm\frac{\nu_1}{2{\cal M}_0}\right)
\\ \displaystyle
\left(\epsilon_0=6G_0,\;\;\epsilon_1=6G_1\pm\frac{\nu_1}{2{\cal M}_0}
\right)\end{array}\right\},\quad\nu_0=-\frac{3}{5}.
\label{fntnQQ}
\ee
On the contrary, for the $A\A$-channel the demand of finiteness of {\it both}
one-particle spectra at $\Lambda\rightarrow\infty$, {\it independently of}
Eq. (\ref{01}), leads to the relations
\bea
\gamma_{\{-\}}\equiv 1,\;\;{\cal M}^{(\pm)}_{0\{-\}}\equiv
\widetilde{\cal M}_0=\frac{\widetilde{\nu}_0}{2G_0},\;\;\nu_{0\{-\}}\equiv
\widetilde{\nu}_0=\frac{3}{5},\;\; \widetilde{\nu}_1=0,\;\;
\widetilde{\epsilon}_{0,1}=4G_{0,1}.\quad
\label{Acnd}
\eea
The spectra may be written as following:
\bea
E_{\{-\}}(k)=E^{(\pm)}_A(k)=\widetilde{E}_{\left\{\!\!\!\tiny
\begin{array}{c}\A \\ A \end{array}\!\!\!\right\}}(k)=\frac{k^2-
\widetilde{\nu}_2}{2\widetilde{\cal M}_0}\pm (4G_2-\widetilde{\epsilon}_2)
\Longrightarrow\frac{k^2}{2\widetilde{\cal M}_0}+\widetilde{\cal M}_0c^2.\;\;
\label{AE1rn}
\eea
So, they are reduced to the standard form for
$\widetilde{\epsilon}_2=4G_2,\;
\widetilde{\nu}_2=-2\left(\widetilde{\cal M}_0c\right)^2$.

As $\gamma_{1\{-\}}\equiv 0$, a non zero solution, similar to Eqs.
(\ref{2states}), (\ref{phichi}), (\ref{Phil}), and (\ref{T0rnrm}) (without
the restriction (\ref{QQ})), appears only if one discriminates the terms
of subsequent order of formally the same divergences $\MV{k^2}$ in Eq.
(\ref{k2}) and ${\MV{[k^2]}}$ in Eq. (\ref{Jnk2}). These divergences
originate from regularizations of the anticommutator (\ref{conv}) in the
one-particle spectrum and the two-particle interaction kernel (\ref{K2}),
respectively. Their difference reflecting their different physical nature
may be easily treated as a fixed shift of the cut-off
$\Lambda\rightarrow\Lambda+\sigma/3$, manifesting itself in
$\MV{[k^2]}\,\rightarrow\,\MV{k^2}+\sigma\Lambda$ and in Eq. (\ref{01}).
However, such a shift makes the above Goldstone solution (\ref{Gld}) to
break down at any finite $\Lambda$ even for $\PP=0$. Thus, the existence of
the bound (and scattering) states in the $A\A$-channel and in the
$AA$- or $\A\A$-channel, as well as the Goldstone mode imply the mutually
exclusive conditions of fine tuning  (\ref{ft+}), (\ref{01}), (\ref{fntnQQ}),
and (\ref{Acnd}).
That is why in Appendix A we trace the further fate of the Goldstone states
and the derivation of the solution (\ref{T0rnrm}) in the framework of
extension theory by means of the procedure which, in a certain sense, is
equivalent to the divergence manipulations of such kind.

Really, a simple normalization test for the scattering solutions (\ref{Phil})
and (\ref{T0rnrm}) shows the necessity of at least one additional discrete
$q$-depended component for the wave function, with a positive or negative
metric contribution according to the sign of $\Upsilon$. So, strictly
speaking, we deal with a self-adjoint extension of the initial free
Hamiltonian, which is restricted on the appropriate subspace of $L^2$, onto
the extended Hilbert (or Pontryagin) space $L^2\oplus C^1$ \cite{Shond3},
\cite{Fewst}. However, this additional discrete component of the
eigenfunctions only corrects their scalar product. It is completely
defined by the same parameters of the self-adjoint extension but does not
affect the physical meaning of the obtained solution in ordinary space
\cite{YShr}-\cite{VDj}. Besides, it would be inappropriate to associate this
additional components with the additional set of creation-annihilation
operators \cite{Shond3} (see Appendix A).

Another extension appears for the choice of finite ``bare'' mass that is
true only for the B-system and for $(A)$-case of the A-system.
Thus $G_{0,1}=\nu_{0,1}=0$, and Eqs. (\ref{gamma1}) and (\ref{gamma}),
together with the condition (\ref{D02}), lead to the solution coinciding
with the well-known extension in $L^2$ \cite{BrFdd} of the singular operator
from Eq. (\ref{R12}) with $\mu_0\equiv 0$, for which:
\bea
&&\gamma^{(-)}_0=1-(3/4)(3\pm\sqrt{5})<1,\qquad
{\cal M}^{(-)}_0=m(3/4)(3\pm\sqrt{5}),\qquad
\nonumber \\
&& \TT{0}{k}\Bigr|_{\Lambda\rightarrow\infty}=
\frac{-\left(2\pi^2\right)^{-1}}{(b\pm iq)},\;\;
\psi^{(0)}_b(\x)=\frac{\sqrt{8\pi b}}{4\pi}\frac{e^{-br}}{r}, \;\;
(r=|\x|).
\label{FB}
\eea
These expressions may be obtained also for the arbitrary $QQ'$-channels
from the previous solution (\ref{T0rnrm}) at the formal limit
$\Upsilon\rightarrow\infty$, what implies that $\sigma\rightarrow\infty$
as independent cut-off.

\section{Three-Particle Eigenvalue Problems. The $QQQ$- Channel.}

The bound-state wave function of three identical particles $Q=\A,B,C,A$
with total momentum $\PP$ is determined by the corresponding Schr\"odinger
equation with the Hamiltonian (\ref{H+H}), (\ref{aH0}) and (\ref{aHI})
$$\widehat{H}|3,\PP\rangle =M_3(\PP)|3,\PP\rangle,$$
where:
\bea
&&|3,\PP\rangle =
\int d^3q_1 d^3q_2 d^3q_3 D^{(\PP,{\rm J,m})}_{\alpha\beta\gamma}
(\q_1\q_2\q_3)\Qop^{\dagger}_{\alpha}(\q_1)\Qop^{\dagger}_{\beta}(\q_2)
\Qop^{\dagger}_{\gamma}(\q_3)\vac_Q ,
\label{3stat}\\
&&D^{(\PP,{\rm J,m})}_{\alpha\beta\gamma}(\q_1\q_2\q_3)
\equiv\left.\delta(\q_1+\q_2+\q_3-\PP)\,{\cal D}^{(\PP,{\rm J,m})}
_{\alpha\beta\gamma}(\q_1\q_2\q_3)\right|_{\q_1+\q_2+\q_3=\PP}
\nonumber \\
&&=-D^{(\PP,{\rm J,m})}_{\alpha\gamma\beta}(\q_1\q_3\q_2)=
-D^{(\PP,{\rm J,m})}_{\beta\alpha\gamma}(\q_2\q_1\q_3)=
-D^{(\PP,{\rm J,m})}_{\gamma\beta\alpha}(\q_3\q_2\q_1).
\label{prop}\\
&&D^{(\PP,{\rm J,m})}_{\alpha\beta\gamma}(\q_1\q_2\q_3)
\left[\sum_{i=1}^3 E_Q(\q_i)-M_3(\PP)\right]=
\label{3Deq} \\
&&=\int d^3k_1 d^3k_2 d^3k_3\,
D^{(\PP,{\rm J,m})}_{\alpha\beta\gamma}(\ka_1\ka_2\ka_3)\,
{\cal H}(\ka_1 \ka_2 \ka_3|\q_1\q_2\q_3),
\nonumber \\
&&{\cal H}(\ka_1\ka_2\ka_3|\q_1\q_2\q_3) =\frac{\lambda}{2(2mc)^2(2\pi)^3}\;
\delta\Bigl(\sum^3_{i=1}\ka_i-\sum^3_{i=1}\q_i\Bigr)\cdot
\label{3kern} \\
&&\cdot \left\{\sum^3_{1=n\neq j<l}\delta(\ka_n-\q_n)\left[(2mc)^2-
(\PP-\q_n)^2+\left(\frac{\ka_j-\ka_l}{2}+\frac{\q_j-\q_l}{2} \right)^2
\right]\right\}.
\nonumber
\eea
The kernel (\ref{3kern}) obviously reproduces all permutation symmetries and
guarantees for momentum conservation. Therefore, it seems convenient to
simplify the separation of the spin-symmetry structure from the coordinate
wave function for $\PP\neq 0$ by using formal functions of three "dependent"
variables, like
${\cal D}^{(\PP,{\rm J,m})}_{\alpha\beta\gamma}(\q_1\q_2\q_3)$ of Eq.
(\ref{prop}), introducing suitable "form factors"
(further on $E(\ka)=E_Q(\ka)\equiv E_{\{+\}}(k)$):
\bea
&&\left.{\cal K}^{(\PP,{\rm J,m})}_{\alpha\beta\gamma}(\q_1\q_2\q_3)
\right|_{\q_1+\q_2+\q_3=\PP}=
\nonumber \\
&&=\left[\sum_{i=1}^3 E(\q_i)-M_3(\PP)\right]\,\left.
{\cal D}^{(\PP,{\rm J,m})}_{\alpha\beta\gamma}(\q_1\q_2\q_3)
\right|_{\q_1+\q_2+\q_3=\PP}.
\label{3KKK}
\eea
Since the momentum conservation condition is totally symmetrical in $\q_j$,
the spin-symmetry structure of ${\cal K}$ and ${\cal D}$ is the same as the
one of $D$. Let hereafter $\{\ldots\}$ mean symmetrization and $[\ldots]$ --
antisymmetrization over internal variables or indices. Then one has three
types of wave functions and corresponding independent "form factors"
\bea
&& {\cal K}^{(\PP,1/2,{\rm m})}_{(X)\alpha\beta\gamma}
(\q_1\q_2\q_3)=
\Gamma^{1/2,{\rm m}}_{\alpha\{\beta\gamma\}}X(\q_1[\q_2\q_3])+
\nonumber \\
&& + \Gamma^{1/2,{\rm m}}_{\gamma\{\alpha\beta\}}X(\q_3[\q_1\q_2])+
\Gamma^{1/2,{\rm m}}_{\beta\{\gamma\alpha\}}X(\q_2[\q_3\q_1])=
\label{K1/2X} \\
&& =\Gamma^{1/2,{\rm m}}_{\alpha\{\beta\gamma\}}
K^{(\PP)}_X(\{\q_1\q_2\}\q_3)-\Gamma^{1/2,{\rm m}}
_{\gamma\{\alpha\beta\}}K^{(\PP)}_X(\{\q_2\q_3\}\q_1);\;\;
\nonumber \\
&&
K^{(\PP)}_X(\{\q_1\q_2\}\q_3)\equiv X(\q_1[\q_2\q_3])+X(\q_2[\q_1\q_3]);
\nonumber \\
&& {\cal K}^{(\PP,1/2,{\rm m})}_{(Y)\alpha\beta\gamma}
(\q_1\q_2\q_3)=
\Gamma^{1/2,{\rm m}}_{\alpha [\beta\gamma]}Y(\q_1\{\q_2\q_3\})+
\nonumber \\
&& +\Gamma^{1/2,{\rm m}}_{\gamma [\alpha\beta]}Y(\q_3\{\q_1\q_2\})+
\Gamma^{1/2,{\rm m}}_{\beta [\gamma\alpha]}Y(\q_2\{\q_3\q_1\})=
\label{K1/2Y} \\
&& =\Gamma^{1/2,{\rm m}}_{\alpha [\beta\gamma]}
K^{(\PP)}_Y([\q_1\q_2]\q_3)-\Gamma^{1/2,{\rm m}}_
{\gamma [\alpha\beta]}K^{(\PP)}_Y([\q_2\q_3]\q_1),
\nonumber \\
&& K^{(\PP)}_Y([\q_1\q_2]\q_3)\equiv Y(\q_1\{\q_2\q_3\})-Y(\q_2\{\q_1\q_3\}),
\nonumber \\
&&K^{(\PP)}_{X,Y}(\q_1\q_2\q_3)+
\left(\mbox{cyclic permutations }(123)\right)=0,
\nonumber \\
&&{\cal K}^{(\PP,3/2,{\rm m})}_{(Z)\alpha\beta\gamma}(\q_1\q_2\q_3)=
\Gamma^{3/2,{\rm m}}_{\{\alpha\beta\gamma\}}K^{(\PP)}_Z([\q_1\q_2\q_3]).
\label{K3/2}
\eea
Here the following properties of the three-spin-wave functions were used:
\bea
&& \Gamma_{\alpha\beta\gamma}^{1/2,1/2}=
a\delta_{\alpha 2}\delta_{\beta 1} \delta_{\gamma 1} +
b\delta_{\alpha 1}\delta_{\beta 2} \delta_{\gamma 1} +
c\delta_{\alpha 1}\delta_{\beta 1} \delta_{\gamma 2},
\qquad\; a+b+c=0,
\nonumber \\
&&\Gamma_{\alpha\beta\gamma}^{1/2,{\rm m}}+
\Gamma_{\gamma\alpha\beta}^{1/2,{\rm m}}+
\Gamma_{\beta\gamma\alpha}^{1/2,{\rm m}}=0,
\label{gam0} \\
&& \Gamma_{\{\alpha\beta \gamma\}}^{3/2,3/2}=
\delta_{\alpha 1} \delta_{\beta 1} \delta_{\gamma 1},\;\;
\Gamma_{\{\alpha\beta\gamma\}}^{3/2,1/2}=
\delta_{\alpha 2}\delta_{\beta 1}\delta_{\gamma 1}+
\delta_{\alpha 1}\delta_{\beta 2}\delta_{\gamma 1}+
\delta_{\alpha 1}\delta_{\beta 1}\delta_{\gamma 2}.
\nonumber
\eea
To change the projection from m to $-$m it is enough to permute the indices
$1\leftrightarrow 2$. For the case J=1/2 the three-spin-functions with
the definite partial symmetry correspond to the eigenvalue of a definite
spin-permutation operator: $\Sigma_{23}=+1$, $(X)$, $b=c$, $a=-2c $, for
the symmetric function $\Gamma^{1/2,{\rm m}}_{\alpha\{\beta\gamma\}}$;
$\Sigma_{23}=-1$, $(Y)$, $b=-c$, $a=0$, for the antisymmetric one
$\Gamma^{1/2,{\rm m}}_{\alpha [\beta\gamma]}$.
All the "form factors" satisfy the same equation and differ only by the
symmetry type $S=X,Y,Z$:
\bea
K^{(\PP)}_S(\q_1\q_2\q_3)=\!\di\! d^3k_1 d^3k_2 d^3k_3\,
\frac{K^{(\PP)}_S(\ka_1\ka_2\ka_3)}{\sum_{i=1}^3
E(\ka_i)-M_3(\PP)}\,{\cal H}(\ka_1\ka_2\ka_3|\q_1\q_2\q_3).\;\;
\label{KSgen}
\eea
Putting for every term of the kernel (\ref{3kern})
$\ka_j-\ka_l=2\s$, $\ka_j+\ka_l=\rv_n $ one has $\rv_n=\PP-\q_n$ and finds
out the general structure of the "form factors" in Eq. (\ref{KSgen}):
\bea
K^{(\PP)}_S(\q_1\q_2\q_3)=\!\!\!\sum^3_{1=n\neq j<l}\!\!\!
\left[(\q_j-\q_l)\vc_{Sn}(\q_n)+\!A_{Sn}(\q_n)+\!(\q_j-\q_l)^2
B_{Sn}(\q_n)\right],\;
\label{Kstrct} \\
\left.\begin{array}{c} A_{Sn}(\q) \\ B_{Sn}(\q) \\
\vc_{Sn}(\q) \end{array}\right\}=
\frac{\lambda}{2(2mc)^2} \di \frac{d^3s}{(2\pi)^3}
\cdot\frac{K^{(\PP)}_S(\ka_1\ka_2\ka_3)}{E(\ka_n)+E(\ka_j)+E(\ka_l)-
M_3(\PP)}\cdot \qquad\quad
\nonumber \\
\cdot\left\{\begin{array}{c}
(2mc)^2+\s^2-(\PP-\q)^2 \\ 1/4  \\ \s \end{array} \right.,
\qquad\qquad\qquad\qquad\qquad\qquad\quad
\label{ABC}
\eea
where, for $1,2,3=n\neq j\neq l$, $j<l$, one has $\ka_n=\q$,
$\ka_j=(\PP-\q)/{2}+\s\equiv\kap_+$,
$\ka_l=(\PP-\q)/{2}-\s\equiv\kap_-$.
The system of integral equations (\ref{Kstrct}) and (\ref{ABC}) may be
simplified by utilizing the symmetry structure of the functions
$K^{(\PP)}_S$ in Eqs. (\ref{K1/2X}), (\ref{K1/2Y}), and (\ref{K3/2}) in
terms of the S-wave and P-wave Faddeev amplitudes
$Q_{Sn}(\q;\p)\equiv A_{Sn}(\q)+\p^2 B_{Sn}(\q)$ and $\vc_{Sn}(\q)$:
\bea
&& \vc_{Z1}(\q)=-\vc_{Z2}(\q)=\vc_{Z3}(\q)\equiv\vc_{Z}(\q);\quad
Q_{Zn}(\q;\p)= 0;
\label{KZ} \\
&&K^{(\PP)}_Z([\q_1\q_2\q_3])=\vc_Z(\q_1)\cdot (\q_2-\q_3)+
\left(\mbox{cyclic permutations }(123)\right);
\nonumber \\
&& Q_{X1}(\q;\p)=Q_{X2}(\q;\p)\equiv Q_{X}(\q,\p);\;\;
Q_{X3}(\q;\p)=-2Q_{X}(\q;\p);
\nonumber \\
&& \vc_{X1}(\q)=\vc_{X2}(\q)\equiv\vc_{X}(\q);\;\;\vc_{X3}(\q)=0;\;\quad
X(\q_1[\q_2\q_3])=
\nonumber \\
&& =Q_{X}(\q_2;\q_1-\q_3)-Q_{X}(\q_3;\q_1-\q_2)+
\vc_X(\q_1)\cdot (\q_2-\q_3);
\label{KX} \\
&& Q_{Y1}(\q;\p)=-Q_{Y2}(\q;\p)\equiv Q_{Y}(\q;\p);\;\;\;
Q_{Y3}(\q;\p)=0;
\nonumber \\
&& \vc_{Y1}(\q)=-\vc_{Y2}(\q)\equiv\vc_{Y}(\q);\;\;
\vc_{Y3}(\q)=- 2\vc_{Y}(\q);\;\;\;\;Y(\q_1\{\q_2\q_3\})=
\nonumber \\
&&=Q_{Y}(\q_1;\q_2-\q_3)+\vc_Y(\q_2)\cdot (\q_3-\q_1)
+\vc_{Y}(\q_3)\cdot (\q_2-\q_1).
\label{KY}
\eea
Solving now each of the systems (\ref{ABC}) together with (\ref{KZ}),
(\ref{KX}), or (\ref{KY}) as nonhomogeneous algebraic systems, where the
unknown integral terms have to be considered as free members, we arrive at
the following three sets of homogeneous Faddeev integral equations:
\bea
&&\vc_Z(\q)=\di\frac{d^3s}{(2\pi)^3}\cdot \frac{1}{(s^2+\varrho^2)}
\left[\frac{-2\mu_0\;\s}{1-\frac{2}{3}{\cal J}_1(\varrho )} \right]
\vc_Z(\kap_+)\cdot (\q-\kap_-);
\label{CZ}\\
&& Q_{X}(\q;2\rv)=\di\frac{d^3s}{(2\pi)^3}\cdot\frac{1}{(s^2+\varrho^2)}
\left[\frac{-\mu_0\,{\cal O}^{\PP-q}_{\{+\}}(\varrho;\s,\rv)}
{{\cal D}^{\PP-q}_{\{+\}}(\varrho )}\right] \cdot
\label{QX} \\
&&\cdot\Bigl\{Q_{X}(\kap_+;\q-\kap_-)-\vc_X(\kap_+)\cdot(\q-\kap_-)\Bigr\};
\nonumber \\
&& \vc_X(\q)=\di\frac{d^3s}{(2\pi)^3}\cdot \frac{1}{(s^2+\varrho^2)}
\left[\frac{\mu_0\;\s}{1-\frac{2}{3}{\cal J}_1(\varrho)} \right] \cdot
\nonumber \\
&&\cdot
\Bigl\{\vc_X(\kap_+)\cdot (\q-\kap_-)+3Q_{X}(\kap_+;\q-\kap_-)\Bigr\};
\nonumber \\
&& Q_{Y}(\q;2\rv)=\di\frac{d^3s}{(2\pi)^3}\cdot\frac{1}{(s^2+\varrho^2)}
\left[\frac{-\mu_0\,{\cal O}^{\PP-q}_{\{+\}}(\varrho;\s,\rv)}
{{\cal D}^{\PP-q}_{\{+\}}(\varrho) }\right]\cdot
\label{QY} \\
&&\cdot\Bigl\{Q_{Y}(\kap_+;\q-\kap_-)+3\vc_Y(\kap_+)\cdot(\q-\kap_-)\Bigr\}.
\nonumber \\
&& \vc_Y(\q)=\di\frac{d^3s}{(2\pi)^3}\cdot\frac{1}{(s^2+\varrho^2)}
\left[\frac{\mu_0\;\s}{1-\frac{2}{3}{\cal J}_1(\varrho)} \right] \cdot
\nonumber \\
&&\cdot\Bigl\{\vc_Y(\kap_+)\cdot(\q-\kap_-)-Q_{Y}(\kap_+;\q-\kap_-)\Bigr\};
\nonumber
\eea
Herein
\bea
&&{\cal O}^{\PP-q}_{\{\pm\}}(\varrho;\s,\rv)\equiv \gamma \MV{[k^2]}
+Z^{\PP-q}_{\{\pm\}}-(2-\gamma)\varrho^2+
\nonumber \\
&& +(1-\gamma)(s^2+\varrho^2+r^2+\varrho^2)+{\cal J}_0(\varrho)
(s^2+\varrho^2)(r^2+\varrho^2);
\label{calO} \\
&&\varrho^2=\varrho^2(q)\equiv \frac{3}{4}q^2+\frac{\PP^2}{4}-
\frac{(\q\PP)}{2}+\omega^2(\PP);\;\;
\omega^2(\PP)={\cal M}_0\Bigl(3E_0-M_3(\PP)\Bigr);
\nonumber \\
&& E(\q)+E(\kap_+)+E(\kap_-)-M_3(\PP)\equiv \left(s^2+\varrho^2\right)/
{\cal M}_0,
\label{3Erho}
\eea
For finite $\Lambda$ one easily recognizes the interiors of the square
brackets in the kernels of these equations as the exact off-shell
extensions (\ref{our}) of the corresponding half-off-shell two-particle
T-matrices from the l.h.s. of Eqs. (\ref{T0rnrm}) and (\ref{T1}). However,
the renormalized versions of these off-shell T-matrices also coincide
with the respective on-shell ones, given by the r.h.s of Eqs.
(\ref{T0rnrm}) and (\ref{T1}) (see Appendix A).
So, one observes, when $\Lambda\rightarrow\infty$, the restoration of
the Galilei invariance, as in the two-particle case \cite{YdFz},
and comes to further simplifications $\vc_{X,Y,Z}=B_{X,Y}=0$. They lead
to one and the same renormalized equation for the only function of
only one variable that determines in principle the coordinate wave function
of the state with ``isospin" 1/2 independently of its spin symmetry:
\bea
X(\q_1[\q_2\q_3])= A(\q_2)- A(\q_3),\quad Y(\q_1\{\q_2\q_3\})= A(\q_1),
\qquad \varsigma=-1,\quad
\label{XYAA} \\
Q_{X}(\q;2\rv)\Longrightarrow Q_Y(\q;2\rv)\Longrightarrow A(\q)=T(\varrho(q))
\,\frac{\varsigma}{2\pi^2}\di d^3s\,\frac{A(\kap_+)}{(s^2+\varrho^2(q))},
\quad
\label{AQXY} \\
\langle\q|{\cal T}(z)|\ka\rangle=
-\lim_{\Lambda\rightarrow\infty}\widehat{t}(-\varrho^2)=
\frac{T(\varrho)}{2\pi^2}=\frac{\left(2\pi^2\right)^{-1}
\Upsilon}{(\varrho-b)(\varrho+\Upsilon+b)}
\stackrel{(\varrho\rightarrow\infty)}{\longmapsto}
\frac{\Upsilon}{2\pi^2\varrho^2}.\quad
\label{ATgn}
\eea

\section{Three-Particle Eigenvalue Problems. The $\A AA$- Channel.}

The case $\A AA$ (or $A\A\A$) looks more intricate, due to its
lower spin symmetry, but in fact it is similar to the previously
considered one. Therefore, we outline only the main points. Defining the
state wave function and its "form factor" as in Eqs. (\ref{3stat}) and
(\ref{3KKK})
\bea
&& |\widetilde{3},\PP\rangle =
\int d^3q_1 d^3q_2 d^3q_3 \widetilde{D}^{(\PP,{\rm J,m})}_{\alpha\beta\gamma}
(\q_1\q_2\q_3)\epsilon_{\alpha\lambda}
\widetilde{\Aop}^{\dagger}_{\lambda}(\q_1)\Aop^{\dagger}_{\beta}
(\q_2)\Aop^{\dagger}_{\gamma}(\q_3)\vac ,
\nonumber \\
&&\left.\widetilde{\cal K}^{(\PP,{\rm J,m})}_{\alpha\beta\gamma}(\q_1\q_2\q_3)
\right|_{\q_1+\q_2+\q_3=\PP}=
\label{AKKK} \\
&&=\left[\widetilde{E}_{\A}(\q_1)+\widetilde{E}_{A}(\q_2)+ \widetilde{E}_{A}(\q_3)-
\widetilde{M}_3(\PP)\right]\,\left.\widetilde{\cal D}^{(\PP,{\rm J,m})}
_{\alpha\beta\gamma}(\q_1\q_2\q_3)\right|_{\q_1+\q_2+\q_3=\PP},
\nonumber
\eea
with $\widetilde{E}(k)\equiv E_{\{-\}}(k)$ from Eqs. (\ref{AE1rn}) and
(\ref{Acnd}), and using the remaining symmetries
\be
\widetilde{\cal K}^{(\PP,{\rm J,m})}_{\alpha\beta\gamma}(\q_1\q_2\q_3)=-
\widetilde{\cal K}^{(\PP,{\rm J,m})}_{\alpha\gamma\beta}(\q_1\q_3\q_2),
\label{tldK123}
\ee
in the notations of Eq. (\ref{gam0}) one observes the following structure,
instead of Eqs. (\ref{K1/2X}), (\ref{K1/2Y}), and (\ref{K3/2}):
\bea
\widetilde{\cal K}^{(\PP,{\rm 1/2,m})}_{(X)\alpha\beta\gamma}(\q_1\q_2\q_3)=
\Gamma^{1/2,{\rm m}}_{\beta\{\gamma\alpha\}}\overline{K}^{(\PP)}_X
(\q_1\q_2\q_3)-\Gamma^{1/2,{\rm m}}_{\gamma\{\alpha\beta\}}
\overline{K}^{(\PP)}_X(\q_1\q_3\q_2),
\label{KXA} \\
\widetilde{\cal K}^{(\PP,{\rm 1/2,m})}_{(Y)\alpha\beta\gamma}(\q_1\q_2\q_3)=
\Gamma^{1/2,{\rm m}}_{\beta[\gamma\alpha]}\overline{K}^{(\PP)}_Y
(\q_1\q_2\q_3)-\Gamma^{1/2,{\rm m}}_{\gamma [\beta\alpha]}
\overline{K}^{(\PP)}_Y(\q_1\q_3\q_2),
\label{KYA} \\
\widetilde{\cal K}^{(\PP,{\rm 3/2,m})}_{(Z)\alpha\beta\gamma}(\q_1\q_2\q_3)=
\Gamma^{3/2,{\rm m}}_{\{\alpha\beta\gamma\}}
\overline{K}^{(\PP)}_Z(\q_1[\q_2\q_3]).\qquad\qquad\qquad
\label{KZA}
\eea
All "form factors" $\overline{K}_S$, $S=X,Y,Z$ obey again the Eq.
(\ref{KSgen}) with obvious replacements in the kernel (\ref{3kern})
and the denominator (see Eq. (\ref{AKKK})). They reveal the same
structure (\ref{Kstrct}) and take the same general form:
\bea
&&\overline{K}^{(\PP)}_X(\q_1\q_2\q_3)\propto\overline{K}^{(\PP)}_Y
(\q_1\q_2\q_3)=\overline{K}^{(\PP)}(\q_1\q_2\q_3).\qquad\qquad\qquad
\label{KXKYK} \\
&&\overline{K}^{(\PP)}_S(\q_1\q_2\q_3)=\sum^3_{1=n\neq j<l}
\left[\overline{Q}_{Sn}(\q_n;\q_j-\q_l)+\overline{\vc}_{Sn}(\q_n)
\!\cdot\!(\q_j-\q_l)\right],\quad
\nonumber \\
&&\overline{Q}_{Z2}(\q;\p)= -\overline{Q}_{Z3}(\q;\p),\quad
\overline{\vc}_{Z2}(\q)=-\overline{\vc}_{Z3}(\q),\quad
\overline{Q}_{Z1}(\q;\p)= 0.\qquad
\label{KZ231}
\eea
Operating as in the previous section we come to the coupled system of
homogeneous Faddeev integral equations for the amplitudes
$\overline{\vc}_{n}(\q)$ and $\overline{Q}_{n}(\q;\p)$ for any $S$,
in contrast to the previous case:
\bea
&&\overline{Q}_{1}(\q;2\rv)=\di\frac{d^3s}{(2\pi)^3}\cdot\frac{1}
{(s^2+\varrho^2)}\left[\frac{\gamma V^*{\cal O}^{\PP-q}_{\{+\}}
(\varrho;\s,\rv)}{{\cal D}^{\PP-q}_{\{+\}}(\varrho )}\right] \cdot
\label{Q1} \\
&&\cdot\Bigl\{\overline{Q}_{2}(\kap_+;\q-\kap_-)+\overline{Q}_{3}(\kap_+;\q-
\kap_-)+\left[\overline{\vc}_2(\kap_+)+\overline{\vc}_3(\kap_+)\right]
\cdot\left(\q-\kap_-\right)\Bigr\};
\nonumber \\
&&\overline{\vc}_1(\q)=\di\frac{d^3s}{(2\pi)^3}\cdot\frac{1}{(s^2+\varrho^2)}
\left[\frac{\gamma V^*\s}{1-\frac{2}{3}{\cal J}_1(\varrho)}\right]\cdot
\nonumber \\
&&\cdot\Bigl\{\left[\overline{\vc}_2(\kap_+)-\overline{\vc}_3(\kap_+)\right]
\cdot\left(\q-\kap_-\right)+\overline{Q}_{2}(\kap_+;\q-\kap_-)-
\overline{Q}_{3}(\kap_+;\q-\kap_-)\Bigr\};
\nonumber \\
&&\overline{Q}_{2}(\q;2\rv)=\di\frac{d^3s}{(2\pi)^3}\cdot\frac{1}
{(s^2+\varrho^2)}\left[\frac{V^*{\cal O}^{\PP-q}_{\{-\}}(\varrho;\s,\rv)}
{{\cal D}^{\PP-q}_{\{-\}}(\varrho )}\right] \cdot
\label{Q2} \\
&&\cdot\Bigl\{\overline{Q}_{1}(\kap_+;\q-\kap_-)+\overline{Q}_{3}(\kap_+;\q-
\kap_-)+\left[\overline{\vc}_1(\kap_+)-\overline{\vc}_3(\kap_+)\right]
\cdot\left(\q-\kap_-\right)\Bigr\};
\nonumber \\
&&\overline{\vc}_2(\q)=\di\frac{d^3s}{(2\pi)^3}\cdot\frac{1}{(s^2+\varrho^2)}
\left[\frac{V^*\s}{1-\frac{2}{3}{\cal J}_1(\varrho)} \right] \cdot
\nonumber \\
&&\cdot\Bigl\{\left[\overline{\vc}_1(\kap_+)+\overline{\vc}_3(\kap_+)\right]
\cdot\left(\q-\kap_-\right)+\overline{Q}_{1}(\kap_+;\q-\kap_-)-
\overline{Q}_{3}(\kap_+;\q-\kap_-)\Bigr\};
\nonumber \\
&&\overline{Q}_{3}(\q;2\rv)=\di\frac{d^3s}{(2\pi)^3}\cdot\frac{1}
{(s^2+\varrho^2)}\left[\frac{V^*{\cal O}^{\PP-q}_{\{-\}}(\varrho;\s,\rv)}
{{\cal D}^{\PP-q}_{\{-\}}(\varrho )}\right] \cdot
\label{Q3} \\
&&\cdot\Bigl\{\overline{Q}_{1}(\kap_+;\q-\kap_-)+\overline{Q}_{2}(\kap_+;\q-
\kap_-)-\left[\overline{\vc}_1(\kap_+)+\overline{\vc}_2(\kap_+)\right]
\cdot\left(\q-\kap_-\right)\Bigr\};
\nonumber \\
&&\overline{\vc}_3(\q)=\di\frac{d^3s}{(2\pi)^3}\cdot\frac{1}{(s^2+\varrho^2)}
\left[\frac{V^*\s}{1-\frac{2}{3}{\cal J}_1(\varrho)} \right] \cdot
\nonumber \\
&&\cdot\Bigl\{\left[\overline{\vc}_2(\kap_+)-\overline{\vc}_1(\kap_+)\right]
\cdot\left(\q-\kap_-\right)+\overline{Q}_{1}(\kap_+;\q-\kap_-)-
\overline{Q}_{2}(\kap_+;\q-\kap_-)\Bigr\}.
\nonumber
\eea
Here we replaced in the definitions (\ref{DDr}) and (\ref{calO}) the
"inverse propagator" from Eq. (\ref{3Erho}) by the one from Eq. (\ref{AKKK})
omitting the term $(\PP-\q,\s)/m$ vanished with $\Lambda\rightarrow\infty$,
what results in the substitution for $\varrho^2(q)$ of Eq. (\ref{3Erho}):
\[
\omega^2(\PP)\longrightarrow\widetilde{\omega}^2(\PP)=\widetilde{\cal M}_0
\Bigl(\widetilde{E}_{\A0}+2\widetilde{E}_{A0}-\widetilde{M}_3(\PP)\Bigr).
\]
Keeping in mind the conditions (\ref{D02}), (\ref{01}) and (\ref{Acnd}), one
finds the same limit (\ref{ATgn}) for the renormalized S-wave kernel of the
first of the Eqs. (\ref{Q2}) and (\ref{Q3}) at $\Lambda\rightarrow\infty$.
However, for the first of Eq.(\ref{Q1}), as well as for all P-wave kernels
above and here, the limit is zero under these conditions. So,
$\overline{\vc}_{1,2,3}(\q)=\overline{Q}_{1}(\q;\p)=0$, and Eqs. (\ref{Q2})
and (\ref{Q3}) degenerate into a system for the functions of only one
variable $\overline{Q}_{2,3}(\q;2\rv))\Longrightarrow\overline{A}_{2,3}(\q)$.
That means
$\varsigma\overline{A}_{3}(\q)=\overline{A}_{2}(\q)\equiv\overline{A}(\q)$,
returning us virtually to the previous Eq. (\ref{AQXY}) for
$A(\q)\longrightarrow\overline{A}(\q)$ with $\varsigma=\pm 1$.
This equation coincides with the Shondin's equation for three-bosonic case
up to a multiplicative constant $\varsigma/2$ \cite{Shond3}. As shown in
refs. \cite{Shond3} and \cite{MMM}, the asymptotic behavior of our separable
off-shell T-matrix (\ref{ATgn}) provides that we deal with a self-adjoint
three-particle Hamiltonian semi-bounded from below in both cases.
However, the Hamiltonians related to more slowly vanishing T-matrices
for other two-particle extensions (\ref{FB}) are unbounded, manifesting the
``collapse" in the three-particle system under consideration.

The absence of any vector parameters for $\PP=0$ implies that
$\overline{A}(\q)\longrightarrow{\cal A}(q)$ for zero total angular momentum
\cite{Sit}
and Eq. (\ref{AQXY}) is reduced as follows:
\be
q\,{\cal A}(q)=T(\varrho (q))\frac{\varsigma}{\pi}\int^\infty_0 dk\,k\,
{\cal A}(k)\ln\left(\frac{k^2+q^2+kq+\omega^2}{k^2+q^2-kq+\omega^2}\right).
\label{A0}
\ee
A simple analysis, carried out in Appendix B, shows that for the
appropriate conditions the integral operator written here is equivalent
to the symmetrical, quite continuous and positively defined.
Therefore, nontrivial solutions of Eq. (\ref{A0})
occur only if $\varsigma\Upsilon>0$: \\
(I) For $\Upsilon>0,\;\varsigma=1$ there are only states with "isospin"
J=1/2 and symmetric wave functions defined by
Eqs. (\ref{AKKK}), (\ref{KXA}), and (\ref{KYA}):
\be
\overline{K}^{(\PP)}_X(\q_1\q_2\q_3)\propto
\overline{K}^{(\PP)}_Y(\q_1\q_2\q_3)\propto\left(\overline{A}(\q_2)+
\overline{A}(\q_3)\right),\;\;
\overline{K}^{(\PP)}_Z=0.
\label{3bwf}
\ee
(II) For $\Upsilon<0,\;\varsigma=-1$ the both states with J=1/2,\,3/2 and
antisymmetric wave functions defined by Eqs. (\ref{AKKK}), (\ref{KXA}),
(\ref{KYA}), and (\ref{KZA}) are possible for $\A AA$-channel,
\be
\overline{K}^{(\PP)}_X(\q_1\q_2\q_3)\propto
\overline{K}^{(\PP)}_Y(\q_1\q_2\q_3)\propto
\overline{K}^{(\PP)}_Z(\q_1[\q_2\q_3])\propto
\left(\overline{A}(\q_2)-\overline{A}(\q_3)\right),
\label{3bwfs}
\ee
as well as the solution (\ref{XYAA}) for $QQQ$-channel.
For $b=0$ the case (I) occurs only.

%%%%%%%%%%%%%%%%%%%%%%%%%%%%%%%%%%%%%%%%%%%%%%%%%%%%%%%%%%%%%%%%

\section{Conclusions}

Let us summarize the main points of our considerations. We picked out
from the various field operator realizations of the singular Hamiltonian
(\ref{H2}) with rich internal symmetry the only realization with
spontaneous symmetry breaking. Then, we revealed the definite
$\Lambda$-dependence of the ``bare'' mass and the coupling constant keeping
the Galilei invariance of the corresponding exact simple Goldstone solutions.
This dependence, in turn, together with a natural subtraction procedure,
fixed the self-adjoint extensions of the Hamiltonian in the one- and
two-particle sectors; the latter determined the well-defined
three-particle Hamiltonian.

So, in ref. \cite{YdFz} and here we have formulated an unambiguous
renormalization procedure extracting renormalized dynamics from a
"nonrenormalizable" contact four-fermion interaction. This procedure
is self-consistent in every $N$-particle sector. It is closely
connected with the construction of the self-adjoint extension of the
corresponding quantum-mechanical Hamiltonians and with the restoration of
Galilei invariance.

It has been shown that the simple $\Lambda$-cut-off and the natural
subtraction prescriptions with the definite $\Lambda$ dependences of
``bare'' quantities fixed by fine-tuning relations reduce the field
Hamiltonian (\ref{H2}) into a family of self-adjoint semi-bounded
Hamiltonians in one-, two-, and three-particle sectors. The above exact
solutions, correctly defined for scattering and bound states, as well as for
the Goldstone mode, contain a finite set of arbitrary extension parameters
${\cal M}_0, E^{(\pm)}_{A0}, b, \Upsilon$ with clear physical meaning
for all two-particle channels of the A,B,C-systems.

Thus, the developed renormalization procedure may be considered as a
direct generalization to strongly singular point interactions of the
Berezin-Faddeev procedure \cite{BrFdd}, \cite{AlGHH}.
From the point of view of quantum field theory it gives an example of a
nonperturbative renormalization for the four-fermion interaction.
It is interesting to note that the initial two-particle operator (\ref{R12})
is the same as the one of Diejen and Tip \cite{VDj}. At the same time,
Shondin's \cite{Shond3} and Fewster's \cite{Fewst} Hamiltonians may be
considered as the various possible renormalized versions of our
renormalized operator defined by Eqs. (\ref{TUxi}) and (\ref{psi1egn}).

The renormalization procedure with the $\Lambda$-cut-off prescription and
fine-tuning relations on the one hand, and extension theory on the other
hand, maintain the same s-wave two-particle solutions (\ref{T0rnrm}) and
(\ref{ATgn}) from the various points thus supplementing each other.
Nevertheless, the additional
physical conditions are necessary to make a choice among the various
mathematical possibilities. E.g., to have a finite spectra for both
particles and antiparticles together with three-particle bound state it is
necessary to consider the $\A AA$-channel with a two-particle bound state in
the $\A A$-channel only, i.e. the case $\nu_0=3/5$.

It is worth to note that, identifying $A_\alpha$ ($\A_\alpha$)
as a ``constituent light $u$ and $d$ quark (antiquark)'' with the
constituent mass $M_N/3\sim m_{\rho}/2\simeq {\cal M}_0=385$ MeV, one
finds from Eq. (\ref{Gld}) for the Goldstone mass
$m_G=(2/5){\cal M}_0=154$ MeV, which is close to the pion mass
$m_{\pi}=140$ Mev. At the same time, the "spinless $\rho$ and $\omega$--
mesons" with the mass $m_{\rho}$ are the nearest two-particle bound states
with the appropriate quantum numbers $l=0$ and J=1,0 \cite{Terent}, what
implies $\Upsilon\gg b\simeq 0$. So, the parameter $\Upsilon$ is sufficient
to reproduce the "nucleon" mass $M_N$ for the solutions (\ref{3bwf}) and
(\ref{efimsol}) with $k=1$, whereas the solutions with $k>1$ describe
qualitatively the "nucleon $P_{11}$--resonances" \cite{Data}.

%%%%%%%%%%%%%%%%%%%%%%%%%%%%%%%%%%%%%%%%%%%%%%%%%%%%%%%%%%%%
The authors are grateful to A.A. Andrianov, R. Soldati and Yu.G. Shondin
for constructive discussions, to V.B. Belyaev and W. Sandhas for useful
remarks, and to the referees of "Few-Body Systems" for suggesting
improvements of the manuscript.
%%%%%%%%%%%%%%%%%%%%%%%%%%%%%%%%%%%%%%%%%%%%%%%%%%%%%%%%%%%%%

\appendix
\section*{Appendix A: The Goldstone Mode "against" Extension Theory.}

Here it is shown how the extension theory maintains the solution
(\ref{T0rnrm}). According to the general Shondin construction
\cite{Shond} developed for our case in ref. \cite{VDj}, self-adjoint
extensions of any operator of type (\ref{R12}) are generated as
extensions of the Laplace operator $H_0=-\nabla^2_{\rm x}$ from the
subspace of ${\cal H}_0=L^2$ $\langle\Xi_j|\psi\rangle =0$ fixed by
functionals
$\langle\ka|\Xi_j\rangle =\Xi_j(\ka)\in{\cal H}_{-j},\;\Xi_1(\ka)=1,\;
\Xi_2(\ka)=\ka^2$ into Pontryagin space of type
${\cal H}_0\oplus C^1\oplus C^1$ with a restriction onto a positively
defined subspace. The resolvents of all such self-adjoint
extensions are contained in the closure (in the Pontryagin space) of
Krein's formula for the resolvent associated with our rank-2 perturbation
(for s-wave):
\bea
\widehat{\mbox{\bf R}}(z)=R_0(z)-R_0(z)|\Xi_j\rangle\left(\Gamma^{-1}(z)
\right)_{jl}\langle\Xi_l| R_0(z),\;\;\; R_0(z)=\left(H_0-zI\right)^{-1},
\label{Res}\\
\Gamma(z)\equiv{\cal K}^{-1}_0+{\cal R}(z)\Longrightarrow
\Gamma(z,\zeta)\equiv{\cal K}^{-1}+{\cal R}(z)-{\cal R}(\zeta),\quad\;\;
z=-\varrho^2,\;
\label{Gamm}\\
{\cal R}_{jl}(z)\equiv\langle\Xi_j|R_0(z)|\Xi_l\rangle
=\frac{(2\pi)^3}{\gamma V^*}{\cal J}_{j+l-2}(\varrho),\;\;
{\cal K}^{-1}_0=\frac{(2\pi)^3}{\gamma V^*}
\left(\!\begin{array}{cc}0 & -1 \\-1 & Z^{\PP} \end{array}\!\right),
\label{L0}\\
{\cal J}_n(\varrho)\stackrel{def}{=}\gamma V^*
\di\limits^\Lambda\frac{d^3k}{(2\pi)^3}\cdot\frac{\left(k^2\right)^n}
{(k^2+\varrho^2)}=
\gamma\MV{[k^2]^{n-1}}-\,\varrho^2\,{\cal J}_{n-1}(\varrho),\qquad\quad
\label{Jnk2}\\
{\cal J}_0(\varrho)=\frac{\gamma V^*}{2\pi^2}
\left(\Lambda-\varrho\,\arctan\frac{\Lambda}{\varrho}\right)
=\frac{\gamma V^*}{2\pi^2}\left(\Lambda-\frac{\pi}{2}\sqrt{\varrho^2}+
\varrho\,\arctan\frac{\varrho}{\Lambda}\right).\qquad
\label{J0}
\eea
This may be rewritten further, using the identity
$R_0(z)|\Xi_2\rangle =|\Xi_1\rangle+zR_0(z)|\Xi_1\rangle$, as:
\bea
&&\widehat{\mbox{\bf R}}(z)=R_0(z)-R_0(z)|\Xi_1\rangle\left\{\widehat{t}(z)\,
\langle\Xi_1|R_0(z)+\Delta(z)\,\langle\Xi_1|\right\}-
\label{RtD} \\
&&-|\Xi_1\rangle\left\{\Delta^*(z)\,\langle\Xi_1|R_0(z)
+\left(\Gamma^{-1}\right)_{22}\langle\Xi_1|\right\},
\nonumber\\
&&\widehat{t}(z)=\left(\Gamma^{-1}\right)_{11}+
z\left[\left(\Gamma^{-1}\right)_{12}+
\left(\Gamma^{-1}\right)_{21}\right]+z^2\left(\Gamma^{-1}\right)_{22},
\label{tzGm}\\
&&\Delta(z)=\left(\Gamma^{-1}\right)_{12}+z\left(\Gamma^{-1}\right)_{22}.
\nonumber
\eea
The first line of Eq. (\ref{RtD}) on the space ${\cal H}_{-1}$ takes a value
in ${\cal H}_{0}$ only, while the second line belongs to
${\cal H}_{-1}\backslash{\cal H}_{0}$. With the help of recurrence relations
(\ref{Jnk2}) and (\ref{J0}) the first identity (\ref{Gamm}) leads to the
expression (\ref{calO}) for:
\be
\langle \s|\Xi_j\rangle\left(\Gamma^{-1}(-\varrho^2)\right)_{jl}
\langle\Xi_l|\ka\rangle =-\frac{\gamma V^*}{(2\pi)^3}
\left[\frac{{\cal O}^{\PP}_{\{\pm\}}
(\varrho;\s,\ka)}{{\cal D}^{\PP}_{\{\pm\}}(\varrho )}\right],
\label{our}
\ee
what gives
\bea
\widehat{t}(z)=-\frac{\gamma V^*}{(2\pi)^3}\left[\frac{\gamma\MV{[k^2]}+Z^{\PP}
_{\{\pm\}}-(2-\gamma)\varrho^2}{{\cal D}^{\PP}_{\{\pm\}}(\varrho )}\right],
\qquad\qquad\qquad\qquad
\label{OUR} \\
\Delta(z)=-\frac{\gamma V^*}{(2\pi)^3}\left[\frac{1-\gamma}
{{\cal D}^{\PP}_{\{\pm\}}(\varrho )}\right],\quad
\left(\Gamma^{-1}(z)\right)_{22}=-\frac{\gamma V^*}{(2\pi)^3}
\left[\frac{{\cal J}_0(\varrho)}{{\cal D}^{\PP}_{\{\pm\}}(\varrho )}\right],
\label{DlGm}
\eea
where for
\bea
&&{\cal D}^{\PP}_{\{\pm\}}(\varrho)\equiv(1-\gamma)^2-{\cal J}_0(\varrho)
\left[\gamma\MV{[k^2]}+Z_{\{\pm\}}^{\PP}-(2-\gamma)\varrho^2\right],
\label{DDr} \\
&&\mbox{ it is implied that }\;{\cal D}^{\PP}_{\{\pm\}}(b)=0.\;\;
\label{D02}
\eea
After the subtraction ${\cal D}^{\PP}_{\{\pm\}}(\varrho)\rightarrow
{\cal D}^{\PP}_{\{\pm\}}(\varrho)-{\cal D}^{\PP}_{\{\pm\}}(b)$, observing
with the condition (\ref{D02}) and the fine-tuning relations (\ref{01}),
the limit $\Lambda\rightarrow\infty$ for Eq. (\ref{our}), as well as for the
T-matrix $\widehat{t}(z)$ in Eq. (\ref{OUR}), certainly leads to solutions
(\ref{T0rnrm}) and (\ref{ATgn}), while
$\Delta(z)$ and $\left(\Gamma^{-1}(z)\right)_{22}$
vanish. However, Eq. (\ref{gamma}) implies for the case $\{-\}$ that
$\Delta(z)\equiv 0$ already for finite $\Lambda$, leaving the resolvent
(\ref{RtD}) in diagonal form with the last term:
\be
-\langle\x|\Xi_1\rangle\left(\Gamma^{-1}(z)\right)_{22}\langle\Xi_1
|\y\rangle =-V^*\,\frac{\delta_3(\x)\delta_3(\y)}{(b^2_G+z)};\;\;
b^2_G(\PP)\equiv\MV{[k^2]}+Z^{\PP}_{\{-\}}=\sigma\Lambda-\nu_2-\PP^2.
\label{RGG}
\ee
So, besides $R_0(z)$, only this term remains for $\sigma=0$ with finite
$b^2_G(\PP)$ according to Eqs. (\ref{gamma}), (\ref{01}), and (\ref{Acnd}).
The generalized solution (\ref{Gld}),
$\sqrt{V^*(2\pi)^{-3}}\,\langle\x|\Xi_1\rangle =\sqrt{V^*}\,\delta_3(\x)$,
is still the exact "wave function" of the Goldstone states (\ref{egnst}) for
$\PP=0$ at finite $m$ and $\Lambda$ as well. However, its contribution (\ref{RGG})
into the resolvent disappears as $\Lambda\rightarrow\infty$ for arbitrary
$\sigma$.
Thus, for $\sigma\neq 0 $, the described procedure gives in fact the limit
of $\widehat{t}(z)$ of Eq. (\ref{OUR}) only, like the procedure in ref.
\cite{Shond3}.

Krein's formula for a resolvent of an extended operator is essentially
the second identity (\ref{Gamm}), where, by definition, the arbitrary finite
constant Hermitian matrix ${\cal K}^{-1}$ has nothing to do with the "bare"
matrix ${\cal K}^{-1}_0$ in Eq. (\ref{L0}).
%%%%%%%%%%%%%%%%%%%%%%%%%%%%%%%%%%%%%%%%%%%%%%%%%%%%%%%%%%%%%%%%%%%%%%%%%%%%
To make it meaningful, as a first step, the pre-Pontryagin space is
constructed by adding to the ${\cal H}_0$-subspace the "generalized defect
elements": $|\psi\rangle =|\phi_1\rangle+c_0|\chi^{(0)}_1\rangle+c_{-1}
|\chi^{(-1)}_2\rangle$, $c_0, c_{-1}\in C^1$, $\phi_1\in{\cal H}_1$,
$|\chi^{(-n)}_j\rangle=
\left(R_0(\zeta)\right)^{j-n}|\Xi_j\rangle\in{\cal H}_{-n}$,
$j-1\geq n\geq 0$, $j=1,2$, where
$\ldots\subset{\cal H}_{1}\subset{\cal H}_{0}\subset{\cal H}_{-1}\subset
\ldots$ is a subscale \cite{VDj} of the usual Sobolev scale \cite{AlGHH} and
$\zeta=-\mu^2<0$ is an {\it arbitrary} subtraction point. In a next step,
the prescription for their scalar products is introduced, which for divergent
cases $m+n>0$ are equated to elements of an arbitrary Hermitian matrix.
Our definition of them incorporates also Berezin's recipe \cite{Ber} with
arbitrary $\lambda^{(m+n)}_{lj}$. It reads:
\be
\langle\chi^{(-m)}_l|\chi^{(-n)}_j\rangle\Longrightarrow a^{(m+n)}_{lj}=
f.p.\left[\langle\chi^{(-m)}_l|\chi^{(-n)}_j\rangle \right]+
P^{(m+n)}_{lj}\left(\lambda^{(m+n)}_{lj}\right),
\label{hihi}
\ee
with the finite part $(f.p.)$ and the diverging polynomial
$P^{(m+n)}_{lj}(\lambda)$, defined by the conditions:
\bea
P^{(m+n)}_{lj}(0)=0,\;\;
f.p.\left[\langle\chi^{(-m)}_l|\chi^{(-n)}_j\rangle \right]=
\lim_{\Lambda\rightarrow\infty}\left[\int\limits^{\Lambda}d^3k\,
\chi^{*(-m)}_l(\ka)\chi^{(-n)}_j(\ka)-P^{(m+n)}_{lj}(\Lambda)\right].
\nonumber \\
\mbox{One has: }\quad
\langle\chi^{(-1)}_2|\chi^{(-1)}_2\rangle
\Longrightarrow a^{(2)}_{22}=2\pi^2\left[
\frac{2}{3\pi}\left(\lambda^{(2)}_{22}\right)^3-
\frac{4}{\pi}\lambda^{(2)}_{22}\mu^2+\frac{5}{2}\mu^3\right],\quad
\label{a(2)22} \\
\langle\chi^{(0)}_2|\chi^{(-1)}_2\rangle\Longrightarrow a^{(1)}_{22}=
2\pi^2\left[\frac{2}{\pi}\lambda^{(1)}_{22}-\frac{15}{8}\mu\right],\;\;
\langle\chi^{(0)}_1|\chi^{(-1)}_2\rangle\Longrightarrow a^{(1)}_{12}=
2\pi^2\left[\frac{2}{\pi}\lambda^{(1)}_{12}-\frac{3}{2}\mu\right].
\nonumber
\eea
However, according to the ref. \cite{VDj}, the linear dependence between the
states $|\chi^{(-m_j)}_{j}\rangle$ with various $j$, including
$|\Xi_1\rangle$, must be eliminated:
\be
|\chi^{(-1)}_2\rangle-|\Xi_1\rangle =\zeta|\chi^{(0)}_1\rangle,\,
\mbox{ i.e.: } a^{(1)}_{22}-a^{(1)}_{12}=-2\pi^2\ \frac{3}{8}\mu,\;\;
\lambda^{(1)}_{22}=\lambda^{(1)}_{12}\equiv\lambda^{(1)},\;\;
\lambda^{(2)}_{22}\equiv\lambda^{(2)}.
\label{1=2}
\ee
Let the function $(-z)^{1/2}$ being a regular branch in the complex plane
cutted at $z>0$ and real-valued at $z<0$, and let for any integer $n>0$:
\bea
{\cal I}_n(z,\zeta)=
(z-\zeta)^n\di\frac{d^3 k\;\left(k^2\right)^n}
{(k^2-z)(k^2-\zeta)^{n+1}}=2\pi(z-\zeta)^n\di\limits^\infty_0
\frac{d\lambda}{\sqrt{\lambda}(\lambda-z)}\left(\frac{\lambda}
{\lambda-\zeta}\right)^{n+1}\quad
\label{Inzzet} \\
= \frac{2\pi^2(-1)^{n+1}}{(z-\zeta)}\left[(-z)^{n+\frac{1}{2}}-
(-\zeta)^{n+\frac{1}{2}}-\sum^n_{s=1}(\zeta-z)^s
\frac{(-\zeta)^{n+\frac{1}{2}-s}(2n+1)!!}{(2s)!!(2n-2s+1)!!}\right].\quad
\nonumber
\eea
Thus, according to the refs. \cite{Shond} and \cite{VDj}, the resolvent
(\ref{Res}) embedded into Pontryagin space
${\Pi_1={\cal H}_0\oplus C^1\oplus C^1}$ reads:
\bea
\vec{\rm R}^t(z)\bigr|_{\Pi_1}={R}^\infty_0(z)-|F_j(z,\zeta)\rangle
\left(\Gamma^{-1}(z,\zeta)\right)_{jl}\langle F^{\dagger}_l(z,\zeta)|,
\;\;\;\mbox{ where:}\qquad\qquad\quad
\label{RhFF} \\
{R}^\infty_0(z)\left[\begin{array}{c}\phi_0\\ \widetilde{u}\\ u\end{array}\right]
=\left(\begin{array}{ccc} R_0(z) & 0 & R_0(z)|\chi^{(-1)}_2\rangle \\
\langle\chi^{(-1)}_2|R_0(z) & 0 & {\cal I}_2(z,\zeta)(z-\zeta)^{-1}+
a^{(1)}_{22} \\ 0 & 0 & 0 \end{array}\right)
\left[\begin{array}{c}\phi_0\\ \widetilde{u}\\ u\end{array}\right],\qquad
\label{R0infty}\\
\Gamma(z,\zeta)={\cal K}^{-1}\!+(z-\zeta)\!\left(\!\begin{array}{cc}
{\cal I}_0(z,\zeta) & {\cal I}_1(z,\zeta)+a^{(1)}_{12} \\
{\cal I}_1(z,\zeta)+a^{(1)}_{12} &{\cal I}_2(z,\zeta)+
(z-\zeta)a^{(1)}_{22}+a^{(2)}_{22}\end{array}\!\right),\quad
\label{Gzz} \\
|F_1(z,\zeta)\rangle =\left(\!\!\begin{array}{c}R_0(z)|\Xi_{1}\rangle \\
\displaystyle {\cal I}_1(z,\zeta)+a^{(1)}_{12} \\ \displaystyle 0
\end{array}\!\!\right),\quad
|F_2(z,\zeta)\rangle=\left(\!\!\begin{array}{c}(z-\zeta)R_0(z)
|\chi^{(-1)}_2\rangle \\
\displaystyle {\cal I}_2(z,\zeta)+(z-\zeta)a^{(1)}_{22} \\ \displaystyle 1
\end{array}\!\!\right),\quad
\label{F1Izzet} \\
\det\left[\frac{\partial\Gamma(z,\zeta)}{\partial z}\right]_{z=\zeta}=
\frac{\pi^2}{\sqrt{-\zeta}}\,a^{(2)}_{22}-\left(a^{(1)}_{12}\right)^2.
\qquad\qquad\qquad\qquad\quad
\label{Shndcond}
\eea
Introducing instead of the $\lambda^{(1),(2)}$ two another nonzero constants
$C,\Upsilon$:
\bea
C=\mu-\frac{2}{\pi}\lambda^{(1)},\qquad
-\Upsilon C^2=\frac{2}{3\pi}\left(\lambda^{(2)}\right)^3-
\frac{4}{\pi}\mu^2\lambda^{(2)}+\frac{2}{\pi}\mu^2\lambda^{(1)}+\mu^3,\quad
\label{CCC} \\
a^{(1)}_{12}=-2\pi^2\left(C+\frac{\mu}{2}\right),\;\;
a^{(1)}_{22}=-2\pi^2\left(C+\frac{7}{8}\mu\right),\;\;
a^{(2)}_{22}=-2\pi^2\left(\Upsilon C^2-
\mu^2C-\frac{\mu^3}{2}\right),\quad
\label{aaCC}
\eea
one find for $z=-\varrho^2$, $\zeta=-\mu^2$ and  {\it arbitrary} Hermitian
matrix ${\cal K}^{-1}$ with elements $\alpha,\beta,\gamma$:
\bea
{\Gamma_{11}(z,\zeta)}={2\pi^2}\left[\alpha+\mu-\varrho\right],\quad
{\Gamma_{12,21}(z,\zeta)}={2\pi^2}\left[\beta+(\varrho-\mu)
\left(\varrho^2+(\varrho+\mu)C\right)\right],\qquad
\label{Grmu} \\
{\Gamma_{22}(z,\zeta)}={2\pi^2}\left[\gamma-(\varrho-\mu)\left(
\varrho^4+(\varrho+\mu)\left(\varrho^2C-\Upsilon C^2\right)\right)\right],
\qquad
\nonumber
\eea
From the asymptotic behavior of $\widehat{t}(z)\simeq\varrho^{-1}$ of Eq.
(\ref{tzGm}) at $\varrho\rightarrow\infty$ we conclude that the solutions
(\ref{T0rnrm}) and (\ref{ATgn}) have no chance to occur for any finite
$\mu,C,\Upsilon$.
However, extending, in certain sense, the possibility to have various
values of $\lambda^{(m+n)}_{lj}$ \cite{Shond}, one opens the way to
obtain the solutions, different from ref. \cite{VDj}, taking the limit
$C\rightarrow\infty$ for fixed $\varrho,\mu,\Upsilon$,
what directly simulates the shift $\sigma$ of cut-off $\Lambda$ in sec.3
above. Now $\Delta(z,\zeta)$ and $\left(\Gamma^{-1}(z,\zeta)\right)_{22}$
vanish again and Eq. (\ref{tzGm}) gives:
\bea
\lim_{C\rightarrow\infty}\widehat{t}(z)=\!\lim_{C\rightarrow\infty}\!
\left(\Gamma^{-1}(z,\zeta)\right)_{11}\!=t(-\varrho^2)=
\frac{-\Upsilon}{2\pi^2(\varrho-{\rm b}_{+})(\varrho-{\rm b}_{-})}
=\frac{-\Upsilon}{2\pi^2(\varrho-b)(\varrho+b+\Upsilon)},
\nonumber \\
b={\rm b}_{\pm},\quad{\rm b}_{\pm}=-\frac{\Upsilon}{2}\pm
\sqrt{\left(\frac{\Upsilon}{2}\right)^2+\frac{\Upsilon}{\xi}},\quad
\frac{1}{\xi}=\alpha+\mu+\frac{\mu^2}{\Upsilon}=
\frac{b}{\Upsilon}(b+\Upsilon)=-
\frac{{\rm b}_{+}{\rm b}_{-}}{\Upsilon},\quad
\nonumber
\eea
where the $\xi$ is a scattering length.
To construct corresponding reduction of the Pontryagin space, we write
from Eqs. (\ref{Gzz}) and (\ref{F1Izzet}) using Eqs. (\ref{CCC}) and
(\ref{aaCC}) at $C\rightarrow\infty$:
\bea
\Gamma^{-1}_C(z,\zeta)\simeq t(z)\left(\begin{array}{cc} 1 & \displaystyle
-\frac{1}{C\Upsilon} \\ \displaystyle -\frac{1}{C\Upsilon} & \displaystyle
\frac{1}{C^2\Upsilon^2}\left[1+\frac{\Upsilon}{2\pi^2t(z)(\zeta-z)}\right]
\end{array}\right),\qquad\qquad
\label{Gm_1tC} \\
|F^C_{1}(z,\zeta)\rangle\!\simeq\!\left(\!\!\begin{array}{c} R_0(z)
|\Xi_1\rangle \\ \displaystyle -2\pi^2C+O(1) \\ \displaystyle 0 \end{array}
\!\!\right),\;
|F^C_{2}(z,\zeta)\rangle\!\simeq\!\left(\!\!\begin{array}{c}(z-\zeta)R_0(z)
|\chi^{(-1)}_2\rangle \\ \displaystyle -2\pi^2C(z-\zeta)+O(1) \\
\displaystyle 1 \end{array}\!\!\right).\quad
\label{F12C}
\eea
The determinant (\ref{Shndcond}) will be well defined for
$\Upsilon\neq -2\mu$.
The metric of the Pontryagin space may be written with the use of
dilatation $\vec{\rm d}_{1/C}={\rm diag}\left\{\vec{\rm I}_0,1/C,C\right\}$
as follows
\bea
\vec{\widehat{\rm g}}_C=\left(\begin{array}{ccc} \vec{\rm I}_0 & 0 & 0 \\
0 & 0 & 1 \\ 0 & 1 & a^{(2)}_{22}\end{array}\right)\simeq
\vec{\rm d}_{1/C}\vec{\widehat{\rm g}}_1\vec{\rm d}_{1/C},\;\;\;
\vec{\widehat{\rm g}}_1=\left(\begin{array}{ccc} \vec{\rm I}_0 & 0 & 0 \\
0 & 0 & 1 \\ 0 & 1 & -2\pi^2\Upsilon\end{array}\right).\qquad
\label{hGÁÁ}
\eea
The invariant subspaces of our self-adjoint Hamiltonian belong to
the subset $\Pi^C_1\subset\Pi_1$ of $C$-depended elements:
\bea
\left(\Pi^C_1\owns\vec{\psi^C}\right)\!\!\iff\!\!
\left(\left[\vec{\psi^C}\right]=\left[\vec{\rm d}_C\vec{\psi^1}\right]=
\left[\phi_0;C\widetilde{h};(1/C)h\right]^T,\;
\left[\vec{\psi^1}\right]=\left[\phi_0;\widetilde{h};h\right]^T\in\Pi^1_1
\right),\quad
\label{PiC1}
\eea
whose inner product does not depend on $C$:
\bea
\langle\vec{\psi}^C|\vec{\psi}^{\prime C}\rangle_{\Pi^C_1}=
\left[\vec{\psi}^{*C}\right]^T\vec{\widehat{\rm g}}_C\left[\vec{\psi}^{\prime C}
\right]=\left[\vec{\psi}^{*1}\right]^T\vec{\widehat{\rm g}}_1
\left[\vec{\psi}^{\prime 1}\right]=
\langle\vec{\psi}^1|\vec{\psi}^{\prime 1}\rangle_{\Pi^1_1}.\quad
\label{Pi11}
\eea
The space $\Pi^1_1$ becomes an invariant renormalized Pontryagin space.
Indeed, from the Eqs. (\ref{R0infty}) and (\ref{Gm_1tC})-(\ref{PiC1}),
it follows that the restriction $\vec{\rm R}^t_C(z)$ of the resolvent
(\ref{RhFF}) onto $\Pi^C_1$ induces the action on $\Pi^1_1$ of
the renormalized resolvent $\vec{\rm R}^{t}_1(z)$ of the renormalized
self-adjoint operator $T_1$ by the rule:
\bea
\left[\vec{\rm R}^{t}_1(z)|\vec{\psi^1}\rangle\right]_{\Pi^1_1}
=\lim_{C\rightarrow\infty}\vec{d_{1/C}}\left[\vec{\rm R}^{t}_C(z)
|\vec{\psi^C}\rangle\right]_{\Pi^C_1}=\lim_{C\rightarrow\infty}\vec{d_{1/C}}
\left[\vec{\rm R}^{t}_C(z)|\vec{d_{C}}\vec{\psi^1}\rangle\right]=\qquad
\label{RCCC} \\
=\left[\!\begin{array}{c}R_0(z)|\phi_0\rangle\\ 0\\ 0\end{array}\!\right]+
\left[\!\begin{array}{c}-R_0(z)|\Xi_1\rangle\\ 2\pi^2\\1/\Upsilon
\end{array}\!\right]t(z)\left(\langle\Xi_1|R_0(z)|\phi_0\rangle-
\frac{\widetilde{h}}{\Upsilon}\right)+\left[\!\begin{array}{c}0 \\ 0 \\
{\cal U}_\perp(z)\end{array}\!\right],\qquad
\nonumber  \\
\mbox{where }\;{\cal U}_\perp(z)=\left(2\pi^2\Upsilon h-\widetilde{h}\right)
\left[2\pi^2\Upsilon(\zeta-z)\right]^{-1}.\qquad
\nonumber
\eea
Now it is a simple matter to see that this space is divided into
invariant subspaces $\Pi^1_1={\cal H}^1_{\Upsilon}\oplus{\cal H}^1_{\perp}$
under the action of the resolvent (\ref{RCCC}):
\be
\Pi^1_1\owns\vec{\psi^1}=\vec{\psi^1_\Upsilon}\oplus\vec{\psi^1_\perp},\;\;
\left[\vec{\psi^1_\Upsilon}\right]=
\left[\phi_0;\,2\pi^2\Upsilon h;\,h\right]^T\in{\cal H}^1_{\Upsilon},\;\;
\left[\vec{\psi^1_\perp}\right]=\left[0_0;\,0;\,h_\perp\right]^T\in
{\cal H}^1_{\perp}.
\label{psiUp1}
\ee
In turn, the action of $\vec{\rm R}^t_1(z)$ on ${\cal H}^1_{\Upsilon}$
is equivalent to the action of resolvent $\vec{\rm R}^t_{\Upsilon}(z)$
of self-adjoint operator $T_\Upsilon$ on the space
${\cal H}_{\Upsilon}={\cal H}_0\oplus C^1$, with metric
$\vec{\widehat{\rm g}}_{\Upsilon}={\rm diag}\{\vec{\rm I}_0,\,2\pi^2\Upsilon\}$,
$\vec{\psi^1_\Upsilon}\mapsto\vec{\psi_\Upsilon}\in{\cal H}_{\Upsilon}$:
\bea
\vec{\rm R}^{t}_1(z)\longmapsto\vec{\rm R}^t_\Upsilon(z)\oplus
\vec{\rm R}^t_\perp(z),\;\;\;T_1\longmapsto T_\Upsilon\oplus T_\perp,\;\;\;
\vec{\rm R}^{t}_\perp(z)|\vec{\psi^1_\perp}\rangle =
(\zeta-z)^{-1}|\vec{\psi^1_\perp}\rangle,\qquad
\nonumber \\
\vec{\psi_\Upsilon}=
\left[\!\!\begin{array}{c}\phi_0 \\ h\end{array}\!\!\right],\;\;
\vec{\rm R}^t_\Upsilon(z)|\vec{\psi_\Upsilon}\rangle
=\left[\!\!\begin{array}{c}R_0(z)|\phi_0\rangle \\ 0\end{array}\!\!\right]+
\left[\!\!\begin{array}{c}-R_0(z)|\Xi_1\rangle \\ 1/\Upsilon\end{array}
\!\!\right]t(z)\left(\langle\Xi_1|R_0(z)|\phi_0\rangle-2\pi^2 h\right),
\nonumber
\eea
where $\langle\vec{\psi}^1_\perp|\vec{\psi}^1_\perp\rangle_{\Pi^1_1}=
-2\pi^2\Upsilon|h_\perp|^2$. The last expression for resolvent in
${\cal H}_{\Upsilon}$ coincides with the ones obtained in
refs. \cite{Shond3} and \cite{Shond}. Extended renormalized operator is
defined by the rule:
\bea
&&\varphi_0(\x)=\frac{u(r)}{4\pi r},\quad
T_\Upsilon\left(\!\begin{array}{c}\varphi_0(\x)\\
\displaystyle -(1/\Upsilon)u(0)\end{array}\!\right)=
\left(\!\begin{array}{c}-(1/r)\partial^2_r(r\varphi_0(\x))\\
\displaystyle u^{\prime}(0)+(1/\xi)u(0)
\end{array}\!\right),
\label{TUxi} \\
&&(1/r)\partial^2_r(r\varphi_0(\x))=\nabla^2\varphi_0(\x)+u(0)\delta_3(\x);
\quad T_\perp|\vec{\psi^1_\perp}\rangle=-\mu^2 |\vec{\psi^1_\perp}\rangle.
\label{psi1egn}
\eea
For the scattering eigenstate which follows from Eqs. (\ref{RtD}) and
(\ref{tzGm}) in pre-Pontrygin space the generalized defect elements are
combined into the vector
$|\Xi_1\rangle=|\chi^{(-1)}_2\rangle-\zeta|\chi^{(0)}_1\rangle$,
regenerating by this way the Goldstone degree of freedom.
Thus, the function $\langle\x|\Xi_1\rangle =(2\pi)^{3/2}\delta_3(\x)$ is
playing a dual role: as a generalized Goldstone state eigenfunction in
the $\Lambda$-cut-off approach or as a total "defect component" of scattering
(and bound) eigenstates in the extended space of the extension theory.
For the final Pontryagin space $\Pi^1_1$ one can associate again the
Goldstone degree of freedom with the additional eigenvector
$|\vec{\psi^1_\perp}\rangle$ of Eq. (\ref{psi1egn}), identifying its
eigevalue from Eq. (\ref{RGG}) for $\sigma=0$, with the use of
Eq. (\ref{AE1rn}) for nonrelativistic $\PP$:\\
$\mu^2\longmapsto b^2_G(\PP)\Longrightarrow
2\left(\widetilde{\cal M}_0c\right)^2-\PP^2>0$.
Note that this state is positively defined only for $\Upsilon<0$ and in
that case it is incorporated into the physical Hilbert space.

The point is that the Goldstone state, considered as a bound state with
zero binding energy and zero angular momentum, is forbidden as a usual
square-integrable solution of the quantum-mechanical Schrodinger equation
with a short-range potential. That is why the purely quantum-field degree of
freedom "disguises" as an additional discrete dimension of the extended
space.
%%%%%%%%%%%%%%%%%%%%%%%%%%%%%%%%%%%%%%%%%%%%%%%%%%%%%%%%%%%%%%%%%%%

\appendix
\section*{Appendix B: Bound-State Faddeev Equation for Zero Total Angular
Momentum.}

Using the hyperbolic substitution with a natural odd continuation of the
function ($\PP=0$)
\bea
&&\frac{q{\cal A}(q)}{T(\varrho(q))}=\varphi(\vartheta)=
-\varphi(-\vartheta);\;\;\;q=\frac{2\omega}{\sqrt{3}}\,\sinh\vartheta;
\label{subst1} \\
&& \varrho(q)=\sqrt{\frac{3}{4}q^2+\omega^2}=\omega\,\cosh\vartheta;\;\;\;
k=\frac{2\omega}{\sqrt{3}}\,\sinh\tau;\;\;\;\varrho(k)=\omega\,\cosh\tau,
\nonumber
\eea
Eq. (\ref{A0}) may be reduced to the following convenient form:
\bea
&&\varphi(\vartheta)=\frac{2\varsigma\wp}{\pi\sqrt{3}}\di\limits^{\infty}_{-\infty}
d\tau\,W(\cosh\tau)\,\varphi(\tau)\,
\ln\left(\frac{2\cosh(\tau-\vartheta)+1}{2\cosh(\tau-\vartheta)-1}\right);
\nonumber \\
&&W(\cosh\tau)=\frac{\omega}{\wp}\cosh\tau\,T\left(\omega\,\cosh\tau\right).
\label{phieq}
\eea
Here, $W(\cosh\tau)$ is an even function of $\tau$ and $\wp$ is a suitable
positive constant introduced for convenience. Note that the last kernel
has additional eigenfunctions with opposite (even) parity.

According to general restrictions from the two- and three-particle
scattering problems \cite{Merk}, a presence of two-particle bound state
$\varrho=b$ implies that $\omega>b\geq 0$. Therefore, if $\Upsilon+2b>0$,
the function $T(\varrho(q))$ from Eq. (\ref{ATgn}) is finite and tends to
zero fast enough to make the following substitution meaningful
\bea
\vartheta=\vartheta(\eta),\quad \tau=\tau(\sigma);\quad
\varphi(\vartheta)=f(\eta)=-f(-\eta);\quad \infty > \chi >0;
\label{subst2} \\
\sigma(\tau)=-\sigma(-\tau)=\di\limits_{-\infty}^{\tau}d\upsilon\, W(\cosh
\upsilon)-\chi;\qquad 2\chi\equiv\di\limits_{-\infty}^{\infty}d\upsilon\,
W(\cosh \upsilon).
\nonumber
\eea
This is obviously true for arbitrary $T(\varrho (q))$ with the above
properties and transforms Eq. (\ref{phieq}) into the equation with a
symmetric and quite continuous kernel \cite{Reed}
\be
f(\eta)=\frac{2\varsigma\wp}{\pi\sqrt{3}}\di\limits^{\chi}_{-\chi}d\sigma\,
f(\sigma)\,\ln\left[\frac{2\cosh\left(\tau(\sigma)-\vartheta(\eta)\right)+1}
{2\cosh\left(\tau(\sigma)-\vartheta(\eta)\right)-1}\right]\equiv
\frac{2\varsigma\wp}{\pi\sqrt{3}}\left(\widehat{O}f\right)(\eta).
\label{symeq}
\ee
With the usual definition of the scalar product in
$L^2(-\chi,\chi)$ for arbitrary function $f(\eta)$ from this space one has
via a Fourier transformation
\bea
&&\left(\widehat{O}f,f \right)=\di\limits^{\infty}_{-\infty}d\varepsilon
\left|{\cal F}(\varepsilon)\right|^2\,\frac{\sinh(\pi\varepsilon/6)}
{\varepsilon\cosh(\pi\varepsilon/2)}> 0;
\label{LL} \\
&&{\cal F}(\varepsilon)=\di\limits^{\infty}_{-\infty}d\tau\,
e^{i\varepsilon\tau}\,W(\cosh\tau)\,\varphi(\tau)\equiv
\di\limits^{\chi}_{-\chi}d\sigma\,f(\sigma)\,e^{i\varepsilon\tau(\sigma)}.
\nonumber
\eea
Therefore, operator $\widehat{O}$ has only positive nonzero eigenvalues and
the finite trace $tr\langle\widehat{O}\rangle=2\chi\ln 3$.

\begin{figure}[htb]
\centering\mbox{\epsfig{file=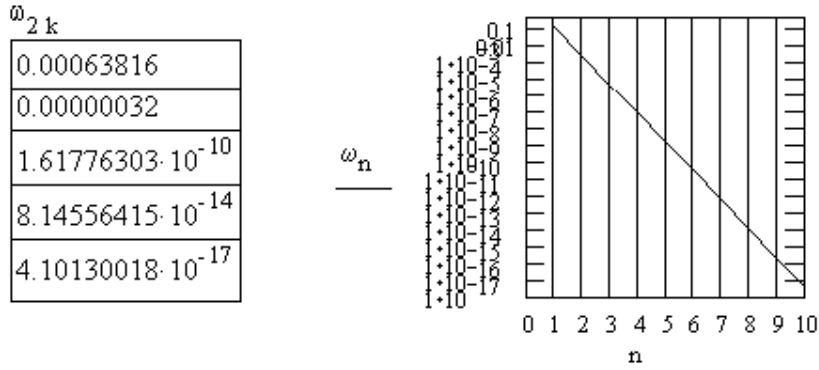,height=5.00cm}}
\protect\caption{Numerical value of energy levels $\omega_{2k}$
and dependence of $\omega_n$ in the units of $\Upsilon$.}
\label{fig}
\end{figure}

\begin{figure}[htb]
\centering\mbox{\epsfig{file=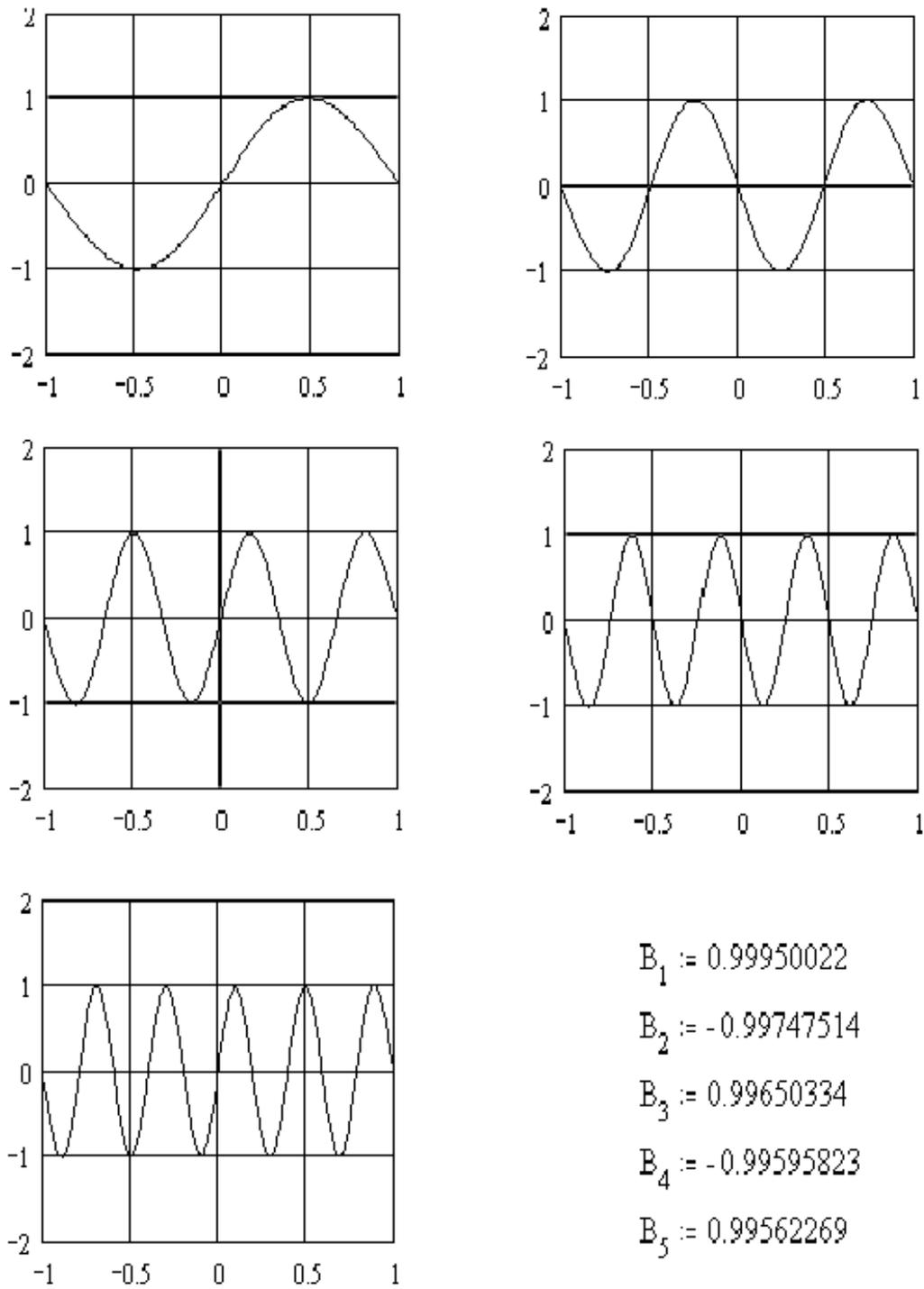,height=20.00cm}}
\protect\caption{Solutions for $k=1,2,3,4,5$ and their Fourier-coefficients
$B_k$.}
\label{graf}
\end{figure}

At last, the simple explicit expressions follow for both $\tau(\sigma)$ and
$\sigma(\tau)$ from Eq. (\ref{subst2}) at $b=0$, $(\Upsilon>0)$, for
$\wp=\coth\chi$, $\omega=\Upsilon/\cosh\chi$:
\be
e^{\tau(\sigma)}=\frac{\sinh\left[(\chi+\sigma)/2\right]}{\sinh\left[
(\chi-\sigma)/2\right]};\qquad
e^{\sigma(\tau)}=\frac{\cosh\left[(\chi+\tau)/2\right]}{\cosh\left[
(\chi-\tau)/2\right]},
\label{nxtsub}
\ee
and similarly for $\vartheta(\eta)$ and $\eta(\vartheta)$.
This allows a direct application of Faddeev's consideration \cite{Merk} to
Eq. (\ref{symeq}) when $\omega\rightarrow 0$, $\chi\rightarrow\infty$.
Thus, $\cosh(\tau-\vartheta)\simeq\cosh(\sigma-\eta)$, and the
seeking of the coefficients $a_m$ of the Fourier expansion
\[
f(\eta)=\sum^{\infty}_{m=-\infty}a_m\,e^{i\pi m\eta/\chi},\;\;
a_m=\sum^{\infty}_{n=-\infty}{\cal C}^{\chi}_{mn}\,a_n,\;\;
{\cal C}^{\chi}_{mn}=\frac{2\varsigma\wp}{\pi\sqrt{3}}\,\frac{1}{2\chi}
\left(\widehat{O}e^{i\pi n\sigma/\chi}, e^{i\pi m\eta/\chi}\right),
\]
leads to the relation of Faddeev type:
\be
1\simeq\frac{4\varsigma }{\sqrt{3}}\,\frac{\sinh(\pi\varepsilon/6)}
{\varepsilon\cosh(\pi\varepsilon/2)},\qquad
\mbox{ for }\;\;\varepsilon\equiv\frac{2\pi m}{2\chi}.
\label{Fadrl}
\ee
It is true for $\varepsilon =\varepsilon_0\simeq 0.4137$, with $\varsigma=1$
only, and gives the asymptotic distribution of Efimov levels and the
respective solutions:
\be
\omega_{2k}=\frac{\Upsilon}{\cosh\chi_k}\simeq\,2\Upsilon
\exp\left\{-\frac{\pi k}{\varepsilon_0}\right\},\quad\;f_{2k}(\eta)\simeq
N_{2k}\sin(\varepsilon_0\eta),\;\quad-\chi_k\leq\eta\leq\chi_k.
\label{efimsol}
\ee
A numerical solution of Eq. (\ref{symeq}) shows that this asymptotic
behaviour in fact starts from the ground state $k=1$ for the interesting odd
solutions $f_{2k}(\eta)$, corresponding to integer $k>0$.
More exactly, for $k=1,2,3,4,5$ one has Eq. (\ref{efimsol}) with
$\chi_k\simeq(k+\delta)\pi/\varepsilon_0$ and $\delta\simeq 0.06006$
(Fig.\ref{fig}). The last value gives also the upper bound of remaining
Fourier-coefficients $|B_{n\neq k}|$ (Fig.\ref{graf}) of the expansion
$f_{2k}(\chi_k y)\simeq\sum^{4k}_{n=0}\left(A_n\cos(\pi ny)+
B_n\sin(\pi ny)\right)$, where $|A_n|\leq 10^{-14}$.
%%%%%%%%%%%%%%%%%%%%%%%%%%%%%%%%%%%%%%%%%%%%%%%%%%%%%%%%%%%%
%---------------------------------------------------------------------------
%----------------------------------------------------------------------
%---------------------------------------------------------------------
%%%%%%%%%%%%%%%%%%%%%%%%%%%%%%%%%%%%%%%%%%%%%%%%%%%%%%%%

\end{document}